\documentclass{svjour3} 
\smartqed  
\usepackage{epsfig}
\usepackage{amsmath}
\usepackage{amssymb}
\usepackage{mathrsfs}
\usepackage{chemarrow}
\usepackage[all]{xy}
\usepackage{times}
\usepackage{comment}
\usepackage{graphicx}

\usepackage{bm}
\usepackage{hyperref}
\hypersetup{
    colorlinks=true,
    linkcolor=blue,
    filecolor=magenta,      
    urlcolor=cyan,
    pdftitle={Sharelatex Example},
}

\def\v1{\mbox{\boldmath$\big|1\big\rangle$}}

\def\vgamma{\mbox{\boldmath$\gamma$}}

\def\mB{\mbox{\boldmath$B$}}
\def\mD{\mbox{\boldmath$D$}}
\def\smD{\mbox{\boldmath{\footnotesize{$D$}}}}

\def\mSigma{\mbox{\boldmath$\Sigma$}}

\def\mJ{{\bf J}}

\def\mV{{\bf V}}

\def\vb{{\bf b}}

\def\vx{{\bf x}}
\def\vy{{\bf y}}

\def\vv{{\bf v}}

\def\rd{{\rm d}}

\newcommand\dbar{\rd\mkern-7mu\mathchar'26\mkern-2mu}

\begin{document}

\title{Mesoscopic and Macroscopic Entropy Balance Equations\\ in a Stochastic Dynamics and Its Deterministic Limit}
\titlerunning{Entropy Balance Equations in Stochastic Dynamics} 

\author{Hong Qian \and Zhongwei Shen
}

\institute{H. Qian \at
Department of Applied Mathematics\\
University of Washington \\
Seattle, WA 98195-3925, U.S.A.\\ 
\email{hqian@u.washington.edu}
\and
Z. Shen \at
Department of Mathematical and Statistical Sciences\\ University of Alberta\\ Edmonton, AB T6G 2G1, Canada\\
\email{zhongwei@ualberta.ca}
}

\dedication{In memory of Professor Min Qian (1927-2019), pioneer in mathematical physics of entropy production}

\maketitle


\begin{abstract}
Entropy, its production, and its change in a dynamical system can be understood from either a fully stochastic dynamic description or a deterministic dynamics exhibiting chaotic behaviors. By taking the former approach based on the general diffusion process with diffusion $\alpha^{-1}\mD(\vx)$ and drift $\vb(\vx)$, where $\alpha$ represents the ``size parameter'' of a system, we show that there are two distinctly different entropy balance equations. One reads $\rd S^{(\alpha)}/\rd t = e^{(\alpha)}_p + Q^{(\alpha)}_{ex}$ for all $\alpha$. Our key result addresses the asymptotic of the entropy production rate $e^{(\alpha)}_p$ and heat exchange rate $Q^{(\alpha)}_{ex}$ up to $O(\tfrac{1}{\alpha})$-corrections as system's size $\alpha\to\infty$. It yields in particular that the ``extensive'', leading $\alpha$-order terms of $e^{(\alpha)}_p$ and $Q^{(\alpha)}_{ex}$ are exactly canceled out. Therefore in the asymptotic limit of $\alpha\to\infty$, there is a second, local entropy balance equation $\rd S/\rd t=\nabla\cdot\vb(\vx(t))+\left(\mD:\mSigma^{-1}\right)(\vx(t))$ on the order of $O(1)$, where $\alpha^{-1}\mD(\vx(t))$ represents the randomness generated in the dynamics usually represented by metric entropy, $\alpha^{-1}\mSigma(\vx(t))$ is the covariance matrix of the local Gaussian description at $\vx(t)$ that is a solution to the ordinary differential equation $\dot{\vx}=\vb(\vx)$ at time $t$, and $\mD:\mSigma^{-1}$ is the Frobenius product of $\mD$ and $\mSigma^{-1}$. This latter equation is akin to the notions of volume-preserving conservative dynamics and entropy production in the deterministic dynamic approach to irreversible thermodynamics {\it \`{a} la} D. Ruelle \cite{ruelle}. Our study follows the rigorous approach and formalism of \cite{jqq}; the mathematical details with sufficient care are given in the appendices.
\end{abstract}


\tableofcontents


\section{Introduction and summary}

There are currently two rather different mathematical frameworks for establishing the concept of entropy in dynamics: Deterministic dynamics \cite{alekseev-yakobson,dorfman,mackey_rmp,ruelle_book} and stochastic processes \cite{jqq,qqt-jsp,schnakenberg,seifert-review}.  It is quite straightforward in the latter to derive the fundamental equation for the balance of entropy, which first appeared in irreversible thermodynamics \cite{prigogine,degroot1962}
\begin{equation}
\label{prigogineEBE}
 \frac{\rd S}{\rd t}  = \frac{\dbar_i S}{\dbar t} + \frac{\dbar_e S}{\dbar t}.
\end{equation}
In the present work we shall denote the rate of entropy production $\dbar_i S/\rd t$ as $e_p$ and the rate of entropy exchange $\dbar_e S/\rd t$ as $Q_{ex}/T$, to emphasize the fact that in general neither quantity on the right hand side of \eqref{prigogineEBE} is a time derivative of any time-dependent thermodynamic state function; both are {\em process dependent} signified by the $\dbar/\dbar t$ \cite{qkkb-review}.  In the deterministic framework an equation like \eqref{prigogineEBE} has been established in the work of D. Ruelle \cite{ruelle} following the  measure-theoretic ``thermodynamic formalism'' \cite{ruelle_book}.  Since Hamiltonian dynamics is Liouvillian volume-preserving, the notion of heat exchange has been widely recognized as the divergence of a vector field in nonlinear dynamics \cite{andrey,dorfman,ramshaw}; see
\cite{cohen_rondoni} and references therein for more extensive literature. The establishment of folding entropy as $e_p$, even its positivity \cite{liao-wang,liu,young1}, is mathematically highly specialized and has been out of the reach for the broader statistical physics community. For example one key result known as Pesin's formula states that the sum of all the positive Lyapunov exponents is equal to the metric entropy of a dynamical system \cite{gaspard_book,qxz,young1}. Because of all this, the relationship between these two approaches to Eq. \eqref{prigogineEBE} and its related decompositions discovered in recent years in stochastic thermodynamics \cite{esposito-prl,gq-pre-10}, has remained unclear.

Deterministic dynamics is the limiting behavior of a stochastic process.  The macroscopic version of Eq. \eqref{prigogineEBE} therefore could be addressed either through (i) deterministic dynamics without ``noise'' as in \cite{ruelle,young1}, or (ii) establishing mesoscopic version of \eqref{prigogineEBE} first as in stochastic thermodynamics followed by the zero noise limit.  By ``mesoscopic'' we mean a description of dynamics that includes fluctuations \cite{qatw_16}, {\em i.e.}, stochastic noise, in contradistinction to ``macroscopic'' behavior with deterministic dynamics. The present paper carries out the limit in (ii); the actual work turns out to be careful computations of the higher order terms in the zero-noise limit, from which a set of results consistent with (i) is revealed.

A deterministic dynamics in (i) is a limit of stochastic descriptions; it can be formulated according to two gross categories: ``averaging in space'' as represented by the $\alpha$ parameter in the present work, or {\em lifting} \cite{wq-jsp-2020} the state space to the space of all paths (see below).  We call the attention that this second perspective is actually already implied in Kolmogorov's mathematical formulation of a stochastic process.  ``Averaging in time'' only yields time-independent equilibrium without a dynamics.  Using finite state discrete time Markov chain as an illustration, we give a brief sketch of this second formulation in Sec. \ref{sec1.1}, taken mainly from \cite{alekseev-yakobson,walters}.  This part is not needed for the main body of the paper, and it could be considered as irrelevant mathematics; but we believe it provides a global understanding of the broader issues on deterministic and stochastic dynamics and their irreversible thermodynamics captured by Eq. \eqref{prigogineEBE} and alike.

The main purpose of this paper is to study the asymptotic of $e_{p}$, $Q_{ex}$, and $\rd S/\rd t$ up to order $O(\tfrac{1}{\alpha})$ as system's size $\alpha\to\infty$. Our key result indicates that both $e_p$ and $Q_{ex}$ are  {\em extensive quantities}, that is, their leading order is $O(\alpha)$, and moreover, their leading terms are exactly the same except having opposite signs. As a result, the macroscopic rate of entropy change $\rd S/\rd t$ is determined. It is not an extensive quantity, but on the order $O(1)$.

The paper is organized as follows: In Sec. \ref{sec1.1}, for the sake of completeness, we provide a very brief summary of the current mathematical formulation of a dynamical system using the simplest finite state discrete time notions.  The important distinction between deterministic and stochastic, or ``path tracking'' vs. ``state tracking'' is outlined. Sec. \ref{sec1.2} introduces the dynamics represented by the Fokker-Planck equation, which is in fact stochastic dynamics with a continuous state space in continuous time and having continuous path.  Our key result regarding the large $\alpha$ asymptotic of $e_{p}$ and $Q_{ex}$ up to order $O(\tfrac{1}{\alpha})$ is derived in Sec. \ref{sec:derivation}. They allow us to connect \eqref{mesoEBE} with the asymptotic of $\rd S/\rd t$ and form a complete understanding of the theory of entropy production. With the mathematical relationship in hand, Sec. \ref{sec:2} further elucidates heuristically the contradistinction between the meaning(s) of entropy production in macroscopic deterministic dynamics and entropy productions in mesoscopic dynamics. Sec. \ref{sec:3} discusses our mathematical results in terms of {\em time scales}, one of the enduring ideas in statistical physics and applied mathematics. The mathematical formulae are familiar \cite{gq-pre-10,gq-pre-16,qian-kinematics}; but the discussion is only possible under Eq. \eqref{key_results}: Dissipative and conservative motions are on the entirely different time scales.  The paper concludes with Sec. \ref{sec:4} which contains a discussion. All the mathematical details for computations are given in the Appendices \ref{app-A}-\ref{app_OU}.

\section{Dynamics: deterministic and stochastic formulations}

\subsection{Stochastic formulation of dynamical systems and its implied topological determinism}
\label{sec1.1}

Kolmogorov's axiomatic theory of stochastic processes with discrete time and the modern ergodic theory of nonlinear dynamics share a common foundation \cite{alekseev-yakobson,walters}.  In its simplest form with a finite state space $\mathcal{S}=\{0,1,\cdots,K-1\}$, one important concept that deserves specific articulation in Kolmogorov's theory is the $\Omega$ space of all elementary events defined by $\Omega=\mathcal{S}^{\mathbb{N}}$; each event is a one-sided infinite sequence of states $\omega=(s_1s_2s_3\cdots)$ with $s_t\in\mathcal{S}$ and $t\in\mathbb{N}$. 
For a given $\omega_0=(s^o_1s^o_2s^o_3\cdots)\in\Omega$, there is a {\em deterministic trajectory in the $\Omega$ space}
\begin{subequations}
\label{infseq}
\begin{equation}\label{infseq1} 
     \omega_0,\cdots,
     \omega_n,\omega_{n+1},\cdots, \, \text{ where } \, 
     \omega_n\in\Omega,
\end{equation} 
that is recursively defined by $\omega_{k+1}=T\omega_k\in\Omega$, where $T$ is called a {\em shift operator} 
\begin{equation}\label{infseq2}
    T\omega \equiv T(s_1s_2s_3\cdots)
    = (s_2s_3s_4\cdots)\in\Omega.
\end{equation}
Moreover, there is a corresponding {\em stochastic trajectory in the state space $\mathcal{S}$} as a coarse-grained description of \eqref{infseq1}
\begin{equation}\label{infseq3}
   s^o_1,\cdots,s^o_n,s^o_{n+1},\cdots, \,
   \text{ where } \,
    s^o_n\in\mathcal{S}, 
\end{equation}
\end{subequations}
Note the states in \eqref{infseq1} are $\omega$'s in $\Omega$ and the states in \eqref{infseq3} are $s$'s in $\mathcal{S}$.

To better illustrate the deterministic dynamics defined by $T$ on $\Omega$, we can employ a topological semi-conjugacy between $T$ on $\Omega$ and the expanding map $E_{K}$ on $[0,1]$ via the representation \cite{walters} 
\begin{equation}
    \omega\equiv (s_1s_2s_3,\cdots)\in\Omega \, \longrightarrow \,  \left( \sum_{n=1}^{\infty}
       \frac{s_n}{K^n}\right) \in 
       [0,1],
\end{equation}
where $E_{k}:[0,1]\to [0,1]$ is defined by
\begin{equation*}
\label{nonlind}
    E_{K}x = Kx \ (\text{mod } 1), \  t\in\mathbb{N}.
\end{equation*}
A deterministic ergodic dynamics means that a single trajectory, consisting of only a countably infinite set of $\omega$ values, is sufficient to statistically represent the entire $[0,1]$ that is a continuum.  We emphasize that there are as many as the entire $[0,1]$ number of different trajectories; all the trajectories corresponding to rational numbers on $[0,1]$ are not ergodic, but they total a set of zero measure and thus are negligible in a statistical sense. 

Heuristically and in a nutshell, each Kolmogorov's elementary event $\omega\in\Omega$ implies an entire trajectory of a deterministic dynamics. Ruelle's thermodynamics \cite{ruelle_book} is based on representations like in \eqref{infseq1} and trajectories determined by $E_{K}$ \cite{ruelle_book}, while stochastic thermodynamics is based on the representation in \eqref{infseq3} and one additional necessary supposition: a probability measure $\mathbb{P}$ on $(\Omega,\mathcal{F})$, under which quantities that involve subsets of $\Omega$ such as 
\begin{equation*}
    \mathbb{P}\{s_1=k_1\}, \, \mathbb{P}\{s_1=k_1,s_2=k_2\}, \text{ and } \, 
    \mathbb{P}\{s_1=k_1,s_2=k_2,s_3=k_3 \}
\end{equation*}
become meaningful, where $\mathcal{F}$ is the cylindrical $\sigma$-algebra of $\Omega$. Simple counting is no longer applicable to quantifying the sizes of random events like $\{s_1=k_1\}$ and $\{s_1=k_1,s_2=k_2\}$, as they are all non-denumerable.\footnote{There is actually a third representation: $s_1^o$, $s_1^os_2^o$, $s_1^os_2^os_3^o$, $\cdots$, which corresponds to a sequence of subsets of $\Omega$ or elements in $\mathcal{F}$, called a {\em filtration}.  In this case, the filtration represents a sequence of increasingly refined sub-$\sigma$-algebra while probability decreases with $n$.  This description is complementary to that in \eqref{infseq2}: the operator $T$ is represented by a ``time mark'': $s_0s_1\cdots\,s_{t-1},s_t\,s_{t+1}\cdots\,\,\to\,\, s_1\cdots\,s_{t-1}\,s_t,s_{t+1}\cdots$. In terms of the filtration, the decreasing size of the subsets corresponds to the increasing information in the sub-$\sigma$-algebra generated by the smaller subsets. In other words, decreasing probability implies decreasing randomness, or fluctuations, and increasing deterministic characteristics.} The probability measure $\mathbb{P}$ provides a more refined description of the shift dynamics $T$ on $\Omega$ than the probability density function on $[0,1]$ for the dynamics $E_{K}$, which is merely a topological representation of the former: There are many different invariant measures of $T$ on $(\Omega,\mathcal{F})$ corresponding to the uniform density on $[0,1]$, defining the unique invariant measure of $E_{K}$ on $([0,1],\mathcal{B}([0,1]))$ in the class of invariant measures absolutely continuous with respect to the Lebesgue measure \cite{young1}, where $\mathcal{B}([0,1])$ is the Borel $\sigma$-algebra of $[0,1]$.

With a given $\mathbb{P}$ on $(\Omega,\mathcal{F})$ that is invariant under $T$, Kolmogorov-Sinai (KS) entropy is the Shannon entropy per step for the entire $\Omega$ \cite{gaspard}.  It can be shown that this KS entropy is never greater than the {\em topological entropy} of $T$ \cite{alekseev-yakobson,ruelle_book}.  The latter being a topological concept, therefore, is intrinsic to the deterministic dynamical system with the path tracking representation in \eqref{infseq}.  It is independent of the probability $\mathbb{P}$; it is non-random.

\subsection{Stochastic dynamics in continuous space and time with continuous path}
\label{sec1.2}

Shift operators for infinite alphabets (state space) and/or continuous time are highly non-trivial mathematical matters. Heuristically, for a continuous time stochastic process on $\mathbb{R}^{N}$, the $\Omega$ is the space of $\mathbb{R}^{N}$-valued functions on $[0,\infty)$; each $\omega\in\Omega$ is a function $\vx(t)\in\mathbb{R}^N$, $t\ge 0$.  The shift operator $T(\tau)$ takes the $\vx(t)$ to $T(\tau)\vx(t)=\vx(t+\tau)$.  It is easy to verify that $T(\tau)$ is a linear operator on the function space; it has various {\em operational calculus representations} \cite{hamermesh}.  One example first employed by Lagrange for smooth $x(t)\in\mathbb{R}$ is via Taylor expansion:
\[
    x(t+\tau)=\sum_{k=0}^{\infty} 
   \frac{\tau^k}{k!} \frac{\rd^k}{\rd t^k}x(t) \equiv
 \left\{\exp\left(\tau\frac{\rd}{\rd t}\right)\right\}x(t).
\]
We shall not follow this line of inquiry except noticing that it has a clear connection to functional analysis and theory of Lie groups. 

As in Eq. \eqref{infseq3}, the stochastic dynamics representation follows the $\vx(t)$ in the state space $\mathbb{R}^N$. The given probability measure $\mathbb{P}$ on $\Omega$ is the Wiener measure, which is also called a white noise.
Under the further assumption that $\vx(t)$ is a continuous function of $t$, the theory of stochastic differential equation is to a $K$-state Markov chain. 

As a mathematical representation of physical movements at such a fundamental level, the very stochastic formulation for $\vx(t)\in\mathbb{R}^N$ already embodies the three most important irreversible dissipative phenomena in statistical physics: diffusion process, heat conduction as energy transfer studied by Fourier, and viscosity via momentum transfer is directly related to diffusion through Einstein's relation. In stochastic thermodynamics, therefore, one asks not how irreversibility arises from deterministic reversible dynamics, rather one studies what reversible dynamics is in stochastic processes \cite{jqq}.  In recent years, the theory of stochastic Markov dynamics has provided a coherent narrative of nonequilibrium thermodynamics at a mesoscopic level; see \cite{qkkb-review,seifert-review} and the references cited within. By ``mesoscopic'', we mean a mechanistic description of dynamical systems of complex individuals in terms of stochastic processes \cite{qian-book}. At the center of this new development is an entropy balance equation, mathematically derivable, in which the notion of {\em entropy production} has been firmly established \cite{schnakenberg,luo,qqt-jsp,SF12,qian_epjst}.  By introducing a ``system's size parameter'' $\alpha$, one can confidently investigate macroscopic thermodynamics as the limit of the mesoscopic description: It has been shown that Gibbs' equilibrium chemical 
thermodynamics of heterogeneous substances is the emergent large deviations theory of a Delbr\"{u}ck-Gillespie process of chemical kinetics \cite{gq-pre-10,gq-pre-16,gq-jsp-17}.  It suggests that ``stochastic kinetics or kinematics dictates energetics'', a saying imitating C. N. Yang's aphorism {\em symmetry dictates interaction} \cite{cnyang}.  See \cite{qian-kinematics,mqw24} for more discussion.

Consider the general diffusion process with continuous state space $\mathbb{R}^N$ and in continuous time, with its Fokker-Planck equation \cite{qqt-jsp,qian_epjst,jqq}
\begin{equation}\label{eq1}
   \frac{\partial f_{\alpha}(\vx,t)}{\partial t}
   = \nabla\cdot \Big( \tfrac{1}{\alpha}\mD(\vx) \nabla f_{\alpha}(\vx,t)-\vb(\vx) f_{\alpha}(\vx,t)\Big),
\end{equation}
where $\alpha$ represents the size of a system, $\vb(\vx)$ is the drift field, $\tfrac{1}{\alpha}\mD(\vx)$ denotes the non-degenerate diffusion. It is understood that $f_{\alpha}(\vx,0)=\delta_{\vx_0}$ with $\vx_0\in\mathbb{R}^{N}$ arbitrarily given. More mathematical details regarding the well-posedness of \eqref{eq1} and basic properties of $f_{\alpha}$ are given in Appendix \ref{app-A}. In addition to many models in physics \cite{HS01}, stochastic formulation of a rapidly stirred nonlinear chemical reaction system follows the {\em chemical master equation}, in which individual molecules are counted one at a time \cite{qian-book}.  In the limit of the system's volume and the number of molecules tend to infinity, their concentrations follow a chemical Langevin equation \cite{gillespie} in which the volume of the chemical reaction vessel is the $\alpha$.

The Shannon entropy functional $S=S[f_{\alpha}]$, which is defined by $$
S=-\int_{\mathbb{R}^N} f_{\alpha}(\vx,t)\ln f_{\alpha}(\vx,t)\rd\vx,
$$ has an instantaneous rate of change \cite{qqt-jsp,jqq,gq-pre-10,VE2010}:
\begin{subequations}
\label{mesoEBE}
\begin{equation}\label{mesoEBE-entropy}
 \frac{\rd S}{\rd t}  = e_p + Q_{ex},
\end{equation}
where the entropy production rate $e_p$ and heat exchange rate 
$Q_{ex}$, measured in units of $k_BT$, are given as
\begin{equation}\label{mesoEBE-entropy-prod-rate}
    e_p = \int_{\mathbb{R}^N} 
       \left[\tfrac{1}{\alpha}\mD(\vx)\nabla f_{\alpha}(\vx,t)-\vb(\vx) f_{\alpha}(\vx,t)\right]\cdot\left[\nabla\ln f_{\alpha}(\vx,t)- \alpha\mD^{-1}(\vx)\vb(\vx) \right]\rd\vx
\end{equation}
and
\begin{equation}\label{mesoEBE-heat-ex-rate}
    Q_{ex} =  \int_{\mathbb{R}^N} 
       \left[\tfrac{1}{\alpha}\mD(\vx)\nabla f_{\alpha}(\vx,t)-\vb(\vx) f_{\alpha}(\vx,t)\right]\cdot\left[\alpha\mD^{-1}(\vx)\vb(\vx) \right]\rd\vx.
\end{equation}
\end{subequations}
While $e_p$ is positive, $Q_{ex}$ can have
either signs.  

If the system is non-driven without external work being done, then the drift $\vb(\vx)$ has a potential $U(\vx)$ such that
\begin{equation}\label{gradient-field}
    \vb(\vx)=-\mD(\vx)\nabla U(\vx).
\end{equation}
It follows that 
\begin{equation*}
\begin{split}
    Q_{ex} &= -\alpha\int_{\mathbb{R}^N} 
       \left[\tfrac{1}{\alpha}\mD(\vx)\nabla f_{\alpha}(\vx,t)-\vb(\vx) f_{\alpha}(\vx,t)\right]\cdot
           \nabla U(\vx) \rd\vx \\
           &= \alpha \frac{\rd}{\rd t}
    \int_{\mathbb{R}^N} f_{\alpha}(\vx,t) U(\vx)\rd\vx
          = \alpha \frac{\rd}{\rd t}\mathbb{E}_{f_{\alpha}}[U]. 
\end{split}   
\end{equation*}
For this class of systems, one can in fact introduce a new functional $F= \alpha\mathbb{E}_{f_{\alpha}}[U]-S$, and then $\rd F/\rd t< 0$.  The potential condition implies that $Q_{ex}$ can be expressed as$\alpha$ times the change of the mean energy $\mathbb{E}_{f_{\alpha}}[U]$. Such a link only exists in equilibrium systems. The stationary state of such a system is an {\em equilibrium} with zero entropy production and minimal $F$.  Eq. \eqref{mesoEBE-entropy} in fact reminds one of the Clausius inequality in thermal physics, where $F$ is called a Helmholtz free energy  \cite{wq-jsp-2020}.


\section{Derivation of the macroscopic entropy balance equation}\label{sec:derivation}

In the limit of system's size $\alpha\to\infty$, the mathematical theory of large deviations (see Appendix \ref{appendix-WKB-transition-prob}) shows that 
\begin{equation}\label{WKB-ansatz}
f_{\alpha}(\vx,t)\sim \left(R_0(\vx,t)+\tfrac{1}{\alpha}R_{1}(\vx,t)\right) e^{-\alpha\varphi(\vx,t)},    
\end{equation}
where the rate function $\varphi(\vx,t)$ solves the Hamilton-Jacobi equation  
\begin{equation*}
-\frac{\partial\varphi(\vx,t)}{\partial t} = \nabla\varphi(\vx,t)\cdot\left(\mD(\vx)\nabla\varphi(\vx,t)+ \vb(\vx)\right), 
\end{equation*}
$R_0(\vx,t)$ solves the linear equation $\mathcal{L}R_{0}(\vx,t)=0$, and $R_{1}(\vx,t)$ solves the non-homogeneous equation $\mathcal{L}R_{1}(\vx,t)=\nabla\cdot(\mD(\vx)\nabla R_0(\vx,t))$. In which, the linear operator $\mathcal{L}$ reads
$$
\mathcal{L}=\partial_{t}+\left(2\mD(\vx)\nabla\varphi(\vx,t)+\vb(\vx)\right)\cdot\nabla +\nabla\cdot\left(\mD(\vx)\nabla\varphi(\vx,t)+\vb(\vx)\right).
$$
Moreover, $\varphi(\vx,t)$ has a global minimum at $\hat{\vx}(t)$
for each $t$, where the function $\hat{\vx}(t)$ is the solution
to the ordinary differential equation (ODE) $\rd\hat{\vx}/\rd t = 
\vb(\hat{\vx})$ with initial value $\hat{\vx}(0)=\vx_0$.  

Applying \eqref{WKB-ansatz}, one can determine the following key results of this paper (see Appendix \ref{app-B-2} for details). Both $e_p$ and $Q_{ex}$ are  {\em extensive quantities}, that is, their leading order is $O(\alpha)$:
\begin{subequations}\label{key_results}
\begin{eqnarray}
    e_p &=& \alpha \vb \cdot  \mD^{-1}\vb  + \mD:\nabla\nabla\varphi +2\nabla\cdot\vb + R  + O\big(\tfrac{1}{\alpha}\big),
\\
	Q_{ex} &=&  - \alpha\vb\cdot \mD^{-1} \vb -\nabla\cdot\vb -R + O\big(\tfrac{1}{\alpha}\big), 
\end{eqnarray}
\end{subequations}
where all the terms on the right-hand side of \eqref{key_results} are calculated at $\vx=\hat{\vx}(t)$, the colon $:$ denotes the Frobenius product of two matrices of the same size, and $R$ is given by
$$
R=\frac{1}{2}\nabla\nabla\big(R_0\vb\cdot\mD^{-1}\vb\big):\mSigma  - \frac{1}{6} \nabla(R_0\vb\cdot\mD^{-1}\vb)\cdot\nabla\nabla\nabla\varphi
              \Theta +\frac{R_{1}\vb\cdot \mD^{-1}\vb}{\sqrt{\det\nabla\nabla\varphi}}.
$$
In which, $\mSigma=\left(\nabla\nabla\varphi\right)^{-1}$ and $\nabla\nabla\nabla\varphi\Theta$ is an $N$-dimensional vector and its $i$-th component is given by
$$
(\nabla\nabla\nabla\varphi\Theta)_{i}=\sum_{j,k,\ell=1}^{N}\partial_{x_{j}x_{k}x_{\ell}}^{3}\varphi\Theta_{ijk\ell},
$$
where 
\begin{equation*}
       \Theta_{ijk\ell}
	 =  \frac{1}{(2\pi)^{N/2}\sqrt{\det\mSigma}}\int_{\mathbb{R}^N}
           y_iy_jy_ky_{\ell} e^{-\frac{\vy\cdot\nabla\nabla\varphi\vy}{2}}\rd\vy.
\end{equation*}

The leading asymptotics of $e_{p}$ and $Q_{ex}$ have been studied and applied to a variety of model systems (see e.g. \cite{FE24}). Our main contribution is the calculation of the $O(1)$ term. We identify a very important fact from \eqref{key_results}: The leading order terms of $e_p$ and $Q_{ex}$ are exactly the same except having opposite signs. As a result, the macroscopic rate of entropy change is not an extensive 
quantity, but it is on the order $O(1)$:
\begin{equation}
   \frac{\rd S}{\rd t}=
  \underbrace{ \nabla\cdot\vb\big(\hat{\vx}(t)\big)
              }_{\text{ local heat exchange}} +
          \underbrace{ \mD\big(\hat{\vx}(t)\big):\nabla\nabla\varphi\big(\hat{\vx}(t),t\big) 
       }_{\text{ local entropy production}} + \  O\big(\tfrac{1}{\alpha}\big).
\label{macroEBE}
\end{equation}
This is expected since $S = -\frac{N}{2}\ln\alpha + O(1)$ (see Appendix \ref{app-B-2}) is not a rapidly oscillating function of $t$ in the limit of system's size $\alpha\to\infty$. The natural logarithmic scaling with the system's size $\alpha$ is a result of taking $\ln$ of $f_{\alpha}(\vx,t)$ whose normalization constant as shown in \eqref{WKB-f-alpha-append} is $\left(\tfrac{\alpha}{2\pi}\right)^{N/2}$ thanks to the non-degeneracy of the rate function $\varphi(\vx,t)$ at its minimum $\hat{\vx}(t)$.

The prefactor in \eqref{WKB-ansatz} only appears within the $R$ in \eqref{key_results}, as a compensation between $e_p$ and $Q_{ex}$ \cite{qian-jjh,qian-eec-01}. While their explicit forms have been difficult to obtain in general (see Appendix \ref{app-B-2}), $R_1=0$ and $R(\vx,t)$ becomes independent of $\vx$ for an Ornstein-Uhlenbeck process with linear drift $\vb(\vx)=\mB\vx$ and constant non-degenerate diffusion $\tfrac{1}{\alpha}\mD$. In fact, according to the precise WKB ansatz \eqref{precise-WKB-ansatz-OU} in this case, $R_{1}=0$ and $\nabla\nabla\nabla\varphi\Theta=0$ as $\varphi$ is quadratic, and therefore, 
$$
R=\frac{1}{2}\nabla\nabla\big(R_0\vb\cdot\mD^{-1}\vb\big):\mSigma=\frac{\left(\mB^{\top}\mD^{-1}\mB\right):\mSigma(t)}{\sqrt{\det\mSigma(t)}},
$$
where $\mSigma(t)=2\int_{0}^{t}e^{\mB(t-s)}\mD e^{\mB^{\top}(t-s)}ds$. The $R$ aside, the emergent macroscopic nonequilibrium thermodynamic structure is completely determined by $\vb(\vx)$, $\mD(\vx)$, and most importantly $\varphi(\hat{\vx},t)$ as an irreversible thermodynamic process.

Even though both Eqs. \eqref{mesoEBE} and \eqref{macroEBE}
are equations for entropy balance, their contradistinction suggests 
subtle but important physics of mesoscopic vs. macroscopic
systems. 

First, in the macroscopic limit, the entropy production rate $e_p$ and
heat exchange rate $Q_{ex}$ become exactly the same; their difference 
yields zero for $\rd S/\rd t$ on the order $O(\alpha)$, that is, the
rate of entropy change is actually very slow.  The macroscopic
heat exchange has the familiar Newtonian frictional form
of ``rate $\vb$ times force $\mD^{-1}\vb$''; it
is the same as entropy production rate, thus non-negative.

	Second, the ``very slow'' rate of entropy change, however,
is a balance between $\nabla\cdot\vb\left(\hat{\vx}(t)\right)$ and 
$\mD\left(\hat{\vx}(t)\right):\nabla\nabla\varphi\left(\hat{\vx}(t),t\right)$.
In the theory of deterministic dynamical systems,
volume preserving dynamics with $\nabla\cdot\vb(\vx)=0$
is considered as ``conservative''.  This is completely
consistent with identifying $\nabla\cdot\vb(\vx)$ as a system's 
energy exchange with the environment, i.e., heat \cite{andrey,ramshaw,ruelle,ruelle99,qwy-pams-19}.  Then in the true stationary state, on the order $O(1)$, there is an exact balance between fluctuations $\mD\left(\hat{\vx}(t)\right):\nabla\nabla\varphi\left(\hat{\vx}(t),t\right)$
and dissipation $-\nabla\cdot\vb\left(\hat{\vx}(t)\right)$.

Third, Eq. \eqref{macroEBE} is actually a generalization of a known result: For an Ornstein-Uhlenbeck process with linear drift $\vb(\vx)=\mB\vx$ and constant non-degenerate diffusion $\tfrac{1}{\alpha}\mD$, the covariance matrix $\tfrac{1}{\alpha}\mSigma(t)$ satisfies 
$\mSigma'(t)=2\mD+\mB\mSigma(t)+\mSigma(t)\mB^{\top}$ 
\cite{qian-prsa-01,cheng_qian_jsp}. Noting that the entropy of a Gaussian
distribution is (see Appendix \ref{app_OU})
$$
S=-\frac{N}{2}\ln\alpha+\frac{N}{2}\ln\left(2\pi e\right)+\frac{1}{2}\ln\Big(\det\mSigma(t)\Big),
$$
one finds from Jacobi's formula that 
$$
\frac{\rd S}{\rd t} =\frac{1}{2}\text{tr}\left[\mSigma^{-1}(t)\mSigma'(t)\right]=\nabla\cdot\vb+\mD:\mSigma^{-1}(t)=\nabla\cdot\vb+\mD:\nabla\nabla\varphi,
$$
where $\varphi(\vx,t)=\tfrac{1}{2}\left(\vx-e^{\mB t}\vx_0\right)\cdot\mSigma^{-1}(t)\left(\vx-e^{\mB t}\vx_0\right)$ is the rate function in this case. The present work establishes a connection between this equation for the variance of a time-dependent Gaussian process, a form of the central limit theory \cite{kurtz}, and the general entropy balance equation in the asymptotic limit
of $\alpha\to\infty$. It shows that the trace of the covariance equation can be interpreted as ``entropy balance''.

Fourth, the local entropy production rate in \eqref{macroEBE}
deserves further elaboration, given in Sec. \ref{sec:2}.


\section{Local entropy production and entropy change}
\label{sec:2}

To illustrate several ideas in stochastic dynamics, let us first consider a Markov chain with a probability distribution $\left\{p_i(t):1\le i\le N\right\}$ at time $t\in\mathbb{N}$ and an one-step transition probability $\left\{P_{ij}:1\le i,j\le N\right\}$. According to the representation in \eqref{infseq1}, which we call ``path tracking'',  
the entropy associated with the probability measure $\mathbb{P}$ for all $\omega$'s, each being a function of time $t\in[0,\infty)$, grows to infinite as $t\to\infty$.  Therefore one defines the rate as $t\to\infty$, also known as entropy per step in the case of discrete time \cite{gaspard}: $-\sum_{i,j=1}^N p_i(t)P_{ij}\ln P_{ij}$.  This expression has also been interpreted as the {\em entropy produced in the dynamic process}, over a single step transition from state $i$ to $j$ being $-\ln P_{ij}$, thus $-\sum_{i,j=1}^N p_i(t)P_{ij}\ln P_{ij}$ being the mean value.

The representation in \eqref{infseq3} which we call ``state tracking'', however, only focuses on the probability distribution among different states at a time $t$, $p_i(t)$.  Therefore it is reasonable to say that not all the ``amount'' of entropy generated from $\omega_t \to \omega_{t+1}$ is being kept in the $p_i(t+1)$: In fact the {\em change in the entropy}
associated with $s_t\to s_{t+1}$ is
\begin{eqnarray}
      -\ln\left(\sum_{k=1}^N p_k(t)P_{kj} \right) + \ln p_i(t),
\label{eq-042}
\end{eqnarray}
with the mean value
\begin{eqnarray*}
  && -\sum_{i,j=1}^N p_i(t)P_{ij} \left[
 \ln\left( \sum_{k=1}^N p_k(t)P_{kj} \right) - \ln p_i(t) \right]
\\
  &=& -\sum_{j=1}^N p_j(t+1)\ln p_j(t+1) +
       \sum_{i=1}^N p_i(t)\ln p_i(t).
\end{eqnarray*}
Therefore the difference between these two differently formulated entropy is \cite{gaspard}
\begin{eqnarray}
	  && \underbrace{ \left\{-\ln P_{ij} \right\} }_{\text{total
  entropy generated}}  -  \underbrace{ \left\{ -\ln\left(\sum_{k=1}^N p_k(t)P_{kj} \right) + \ln p_i(t) \right\} }_{\text{ entropy change
in the system}} = -\ln\left(
        \frac{ p_i(t)P_{ij} }{ \sum_{k=1}^N p_k(t)P_{kj} }\right)
\nonumber\\
	&=& \underbrace{ -\ln\left( \Pr\left\{ X(t)=i \big| X(t+1)=j\right\} \right) 
  }_{\text{entropy discarded is associated with
   uncertainty in the past} } \geq 0.
\label{eq-043}
\end{eqnarray}
The quantity $-\ln P_{ij}$ is the entropy change in a path tracking representation of a Markov chain \eqref{infseq1}, Eq. \eqref{eq-042} is the entropy change in the state tracking representation \eqref{infseq3} of the same Markov chain. The latter representation is a projection of the former. The difference given in \eqref{eq-043} is indeed related to 
the folding entropy in \cite{ruelle,liu,liao-wang}. Its interpretation is intimately related to A. Ben-Naim's notion of entropy of assimilation \cite{ben-naim}.  {\em The difference between the uncertainty
in the future and the uncertainty in the past is the entropy change in the system}:  One verifies a connection between the entropy change in a system according to the state tracking and past-future asymmetry in the path tracking.

Parallel to the above discussion on a discrete-time and discrete-space Markov chain, for a continuous-time and continuous-space Gaussian process with variance $\sigma^2$ at time $t$ and a Brownian step from $t$ to $t+\tau$ with a diffusion coefficient $D$, the probability distribution at $t+\tau$ is again Gaussian with variance $\sigma^2+2D\tau$. The instantaneous rate of entropy change in the continuous system, corresponding to \eqref{eq-042}, then is 
\begin{equation}
  \lim_{\tau\to 0}
   \frac{  \ln\sqrt{2\pi e(\sigma^2+2D\tau)} - 
 \ln \sqrt{2\pi e\sigma^2}  }{\tau} = \frac{D}{\sigma^2}.
\label{eq11}
\end{equation}
Applying the result in \eqref{eq11} to the local
entropy production rate in \eqref{macroEBE}, $\nabla\nabla\varphi\left(\hat{\vx}(t),t\right)$ is the inverse of the local Gaussian variance at $\hat{\vx}(t)$
at time $t$. Therefore, $\mD(\hat{\vx}(t)):\nabla\nabla\varphi\left(\hat{\vx}(t),t\right)$
is the local entropy production at time $t$ that is kept by the system, due to the randomness in the dynamics. See \cite{qian-kinematics} for a more thorough mathematical 
analysis of the diffusion process in the limit of $\alpha\to\infty$.

The local entropy production is no longer a characterization 
of the violation of detailed balance or time irreversibility, two essential concerns in stochastic thermodynamics. Rather, it is a characterization of the randomness generated 
by the dynamics, $\mD$, and the uncertainties in the system, 
$\mSigma^{-1}=\nabla\nabla\varphi$.

We use the term ``local'' to signify the following fact:  The mean value of stochastic thermodynamic quantities such as $e_p$, $Q_{ex}$, as well as $F$ and $Q_{hk}$ below, are all integrals of the distribution function $f_{\alpha}(\vx,t)$ over the entire state space $\mathbb{R}^{N}$.  However in the limit of $\alpha\to\infty$ the $f_{\alpha}(\vx,t)\to \delta\big(\vx-\hat{\vx}(t)\big)$.  The corresponding macroscopic thermodynamic quantities are now functions of $\hat{\vx}(t)$ which is a single point in $\mathbb{R}^N$ at a time. Therefore in contrast to the former, the latter is defined locally in the state space.

\section{Time scales in the nonequilibrium thermodynamic description of stochastic dynamics} 
\label{sec:3}

\subsection{Free energy balance equation}

For stochastic mechanical systems in contact with a heat bath, entropy is not the appropriate equilibrium thermodynamic potential, free energy is.  In classical thermodynamics, the 
difference between entropy and free energy is the mean internal energy. This is reflected in the generalized free energy (also called non-adiabatic entropy \cite{seifert-review}, Kullback–Leibler divergence or relative entropy \cite{BL55,CT05}) $F=F[f_{\alpha}]$ defined by
\begin{equation*}
            F = \int_{\mathbb{R}^N}
                  f_{\alpha}(\vx,t)\ln\frac{ f_{\alpha}(\vx,t) }{\pi_{\alpha}(\vx)}\rd\vx,
\end{equation*}
where $\pi_{\alpha}(\vx)$ is the positive stationary solution to the
Fokker-Planck equation \eqref{eq1} and satisfies $\int_{\mathbb{R}^{N}}\pi_{\alpha}(\vx)\rd\vx=1$. For fixed system's size $\alpha$, $F\to0$ as $t\to\infty$, and therefore, it describes the relaxation of $f_{\alpha}(\vx,t)$ to the steady state $\pi_{\alpha}(\vx)$ as $t\to\infty$.

For equilibrium systems, that is, $\vb(\vx)$ satisfies \eqref{gradient-field}, we have $\pi_{\alpha}(\vx)=\tfrac{1}{K_{\alpha}}e^{-\alpha U(\vx)}$ with $K_{\alpha}=\int_{\mathbb{R}^{N}}e^{-\alpha U(\vx)}\rd\vx$. As a result, $F=\alpha\mathbb{E}_{f_{\alpha}}[U]-S+\ln K_{\alpha}$. Introducing $\widetilde{U}=\alpha^{-1}\ln K_{\alpha}$ yields $F=\alpha\mathbb{E}_{f_{\alpha}}[\widetilde{U}]-S$, which is analogous to the Helmholtz free energy \cite{esposito-prl}.

The functional $F$ also has a balance equation of its own \cite{HS01,gq-pre-10,esposito-prl}:
\begin{equation}\label{mesoFBE-1}
 \frac{\rd F}{\rd t} = -e_p + Q_{hk},
\end{equation}
where the house-keeping heat rate $Q_{hk}$ (also known as the adiabatic entropy production rate) is
\begin{equation*}\label{mesoFBE-2}
	 Q_{hk} =  \int_{\mathbb{R}^N} 
       \left[\tfrac{1}{\alpha}\mD(\vx)\nabla f_{\alpha}(\vx,t)-\vb(\vx)f_{\alpha}(\vx,t)\right]\cdot\left[
          \nabla\ln\pi_{\alpha}(\vx) - \alpha\mD^{-1}(\vx)\vb(\vx) \right]\rd\vx.
\end{equation*}
It is also a non-negative quantity, and is actually positive unless $\vb(\vx)$ is a gradient field \eqref{gradient-field} (see Appendix \ref{appendix-free-energy-balance-eqn}). The non-negativity of both $e_p$ and $Q_{hk}$ suggests that they can be identified as ``sink'' and ``source'' of the generalized free energy $F$ in a 
nonlinear stochastic dynamical system. Moreover, the "sink" is stronger than the "source", resulting in the nice Lyapunov property $\rd F/\rd t<0$. This is a well-known fact; we provide the details in Appendix \ref{appendix-free-energy-balance-eqn} for completeness.

	In the limit of size parameter $\alpha\to\infty$, both
$F$ and $Q_{hk}$ are extensive quantities on the order of
$O(\alpha)$ (see Appendix \ref{appendix-asymptotic-generalized-free-energy} for details): 
\begin{subequations}\label{free-energy-asymptotic}
\begin{eqnarray}
    F &=& \alpha\varphi^{ss}\big(\hat{\vx}(t)\big)+\mathcal{C}_{\alpha}
    + O(1),
\\
	\frac{\rd F}{\rd t} &=& -\alpha \left[
   \nabla\varphi^{ss}\big(\hat{\vx}(t)\big)\cdot\mD\big(\hat{\vx}(t)\big)  \nabla\varphi^{ss}\big(\hat{\vx}(t)\big) \right] + O(1), 
\\
     Q_{hk} &=& \alpha \left[
   \vgamma\big(\hat{\vx}(t)\big)\cdot\mD^{-1}\big(\hat{\vx}(t)\big)  \vgamma\big(\hat{\vx}(t)\big) \right] + O(1),
\end{eqnarray}
\end{subequations}
where $\varphi^{ss}(\vx)$ is the leading-order 
exponent of $\pi_{\alpha}(\vx) \sim e^{-\alpha\varphi^{ss}(\vx)}$, $\mathcal{C}_{\alpha}$ is a constant satisfying $\lim_{\alpha\to\infty}\tfrac{1}{\alpha}\ln\left(|\mathcal{C}_{\alpha}|+1\right)=0$,
and $\vgamma(\vx)=\vb(\vx)+\mD(\vx)\nabla\varphi^{ss}(\vx)$. In the case that the ODE $\dot{\vx}=\vb(\vx)$ admit a globally asymptotically stable and non-degenerate equilibrium, there holds $\mathcal{C}_{\alpha}=0$. The leading terms in the asymptotic \eqref{free-energy-asymptotic} have been discussed in \cite{qian-kinematics,FE22} with applications to the emergent second law of thermodynamics. They are also used in \cite{SFEF24} to bound asymptotic escape rates from metastable states.

One of important results from the theory of large deviations \cite{fw} is that 
\begin{equation}\label{stationary-HJE}
    \vgamma(\vx)\cdot\nabla\varphi^{ss}(\vx)=0.
\end{equation} 
In fact, Eq. \eqref{mesoFBE-1}, on the macroscopic
scale, becomes a Pythagorean triangle equality:
\begin{equation*}
     \left\|\mD\big(\hat{\vx}(t)\big)\nabla\varphi^{ss}\big(\hat{\vx}(t)\big)\right\|^2_{\smD^{-1}\left(\hat{\vx}(t)\right)}  +  
\big\|\vgamma\big(\hat{\vx}(t)\big)\big\|^2_{\smD^{-1}\left(\hat{\vx}(t)\right)}
= \big\|\vb\big(\hat{\vx}(t)\big)\big\|^2_{\smD^{-1}\left(\hat{\vx}(t)\right)},   
\end{equation*}
where the norm of a vector $\vv$ is defined as 
$\|\vv\|_{\smD^{-1}(\vx)} = \sqrt{\vv\cdot\mD^{-1}(\vx)\vv}$ \cite{qian-kinematics}.

\subsection{Time scales, short-time and long-time perspectives}

Among all the thermodynamic quantities we discussed
above, macroscopic $F$, $Q_{hk}$, and $\rd F/\rd t$ are 
defined via $\varphi^{ss}(\vx)$, while macroscopic
$e_p$ and $Q_{ex}$, and local heat exchange rate
are functions of $\vx$ via $\mD(\vx)$ and 
$\vb(\vx)$.  $\rd S/\rd t$ and its corresponding
local entropy production rate, however, are defined via
$\varphi(\vx,t)$, which is dependent upon the choice of
$\varphi(\vx,0)$, e.g., initial fluctuations.  
The $\varphi^{ss}(\vx)$ plays the role
of a macroscopic potential energy function.  In the 
macroscopic, deterministic limit, free energy and its
rate of change are dominated by this energy function
$\varphi^{ss}(\vx)$.

	The fact that $\rd F/\rd t \sim O(\alpha)$ and 
$\rd S/\rd t \sim O(1)$ tells us that in time-dependent 
nonequilibrium thermodynamics, free energy relaxation is
fast, while entropy change is slow. This is a point that has escaped general attention in the past discussion on nonequilibrium thermodynamics.  Through $\varphi^{ss}(\vx)$,
free energy is a ``global'' characterization of 
the nonlinear stochastic dynamics, while entropy dynamics
is only local.

The physical meaning of $\varphi^{ss}\big(\hat{\vx}(t)\big)$ 
and $\varphi\big(\hat{\vx}(t),t\big)$, where $\hat{\vx}(t)$ being
the solution to the deterministic motion as the solution to
$\rd\hat{\vx}/\rd t = \vb (\hat{\vx})$, deserves further 
discussion: $\nabla\varphi\big(\hat{\vx}(t),t\big)=0$ for all
time $t$. The matrix $\mSigma\left(\hat{\vx}(t),t\right)=\left(\nabla\nabla\varphi\right)^{-1}\left(\hat{\vx}(t),t\right)$, as a function of $\hat{\vx}(t)$, 
is the local fluctuation in the asymptotic limit of
$\alpha$ being very large but not infinite.  It is usually called 
time-dependent {\em fluctuations}.  There is a time-dependent
Gaussian process that describes this regime, following
the theory of van Kampen's system-size expansion, or 
Keizer's nonequilibrium statistical thermodynamics 
\cite{keizer-book}. $\nabla\varphi^{ss}\big(\hat{\vx}(t)\big)$,
on the other hand, tells eventually in a very long time scale,
the local probability and its gradient at $\hat{\vx}(t)$.  It 
also provide a definitive statement that
$\rd\varphi^{ss}\big(\hat{\vx}(t)\big)/\rd t=$
$\vb\big(\hat{\vx}(t)\big)\cdot\nabla\varphi^{ss}\big(\hat{\vx}(t)\big)$
$\le 0$ is never positive, thanks to Eq. \eqref{stationary-HJE}.  $\varphi^{ss}(\vx)$ is a landscape for the dynamics with permanence, irrespective of the
initial $\varphi(\vx,0)$.  

All this discussion reflects a ``competition'' between the two limits $\alpha\to\infty$ and $t\to\infty$ in an ergodic system, as first understood in the theory of equilibrium phase transition: With larger and larger $\alpha$, a trajectory with less stochasticity will take longer and longer time to visit and re-visit all the space states. When $\alpha=\infty$, ergodicity is broken in nonlinear deterministic dynamics with multiple attractors.  However, no matter how large $\alpha$ is, as long as $\alpha<\infty$, there will be enough time for the dynamics to cover the whole state space as $t\to\infty$.  See \cite{qatw_16} for an extensive discussion on this issue.


\section{Discussion}
\label{sec:4}

In current textbooks on equilibrium statistical mechanics, thermodynamic limit is rigorously defined as a system's
size, $\alpha$, tending to infinity.  There is no statement
on the time scales of the rate of changes in thermodynamic 
quantities, particularly their dependency upon the system's size $\alpha$.  Time scale(s), however, is certainly central to physics.  We discover that there is a deep relation between the $\alpha\to\infty$ and time scales of an entropy 
balance and for a free energy balance:
The former is on the order of $O(1)$ but the latter is on the 
order $O(\alpha)$.  

There have been two approaches to statistical thermodynamics, 
one based on classical mechanics originated by L. Boltzmann, 
and one based on probability originated by J. W. Gibbs
\cite{lebowitz99}.  Development in nonlinear dynamical systems based on chaotic hypothesis \cite{gallavotti} 
and Sinai-Ruelle-Bowen measure \cite{qxz,young1,young2} is 
the continuation of the former \cite{ruelle99,dorfman}, while 
the stochastic thermodynamics \cite{qkkb-review,seifert-review} 
and in-depth explorations of the theory of probability were the further development of the latter \cite{jqq,gq-pre-10,gq-pre-16,wq-jsp-2020,cqz-2020}.  The present result provides a natural logic bridge between the entropy balance equations, as the fundamental of nonequilibrium thermodynamics, that emerge in these two approaches. If one identifies $\alpha=\epsilon^{-1}$ where 
$\epsilon$ being the size of a Markov partition for a 
deterministic dynamical system, then taking the limit
of $\alpha\to\infty$ is consistent with the modern
treatment in terms of a generating partition which gives
rise to Kolmogorov-Sinai metric entropy and the 
nonequilibrium thermodynamics {\it \`{a} la} D. Ruelle
\cite{dorfman}.

Our present result might also have implications to equilibrium
thermodynamic analysis in which researchers are routinely
partitioning the energetic and entropic contributions to total free 
energy change via van't Hoff method \cite{schellman}.  
A compensation between the 
entropy and energy changes has been extensively discussed
in the past \cite{qian-jjh,qian-eec-01}.  
With $\Delta S = 0$ on $O(\alpha)$ in
$\Delta t$ time, it is tempting to interpret $e_p\Delta t$ 
and $Q_{ex}\Delta t$ as entropy change and energy change 
in a quasi-stationary process.

\section*{Acknowledgement}

We thank Yu-Chen Cheng, Alex Grigo, Yves Pomeau, Shirou Wang, Lai-Sang Young, and particularly Krzysztof Burdzy for helpful discussions. Z. S. was partially supported by a start-up grant from the University of Alberta, NSERC RGPIN-2018-04371, and NSERC RGPIN-2024-04938. We are grateful to the referees for carefully reading the manuscript and providing many constructive critiques and helpful suggestions, which have led to a significant improvement of the paper.

\section*{Conflict of interest statement}

The authors declare that they have no conflict of interest.

\section*{Data availability statement}

This paper has no associated data.


\appendix

\section{Fokker-Planck equation and WKB ansatz}
\label{app-A}

Consider the following Fokker-Planck equation
\begin{subequations}\label{fp-ic-eqn}
\begin{equation}\label{fp-eqn}
   \partial_{t} f_{\alpha}
   = \nabla\cdot \left( \tfrac{1}{\alpha}\mD\nabla f_{\alpha}-\vb f_{\alpha}\right)\quad\text{in}\quad \mathbb{R}^{N},
\end{equation}
\begin{equation}\label{ic-eqn}
    f_{\alpha}(\cdot,0)=\delta_{\vx_0}
\end{equation}
\end{subequations}
where $\alpha$ represents the size of a system, $\vx_{0}\in\mathbb{R}^{N}$, $\delta_{\vx_0}$ denotes the Dirac measure at $\vx_0$, the diffusion matrix $\mD=\mD(\vx)$ is symmetric positive definite and sufficiently smooth, and the vector/drift field $\vb=\vb(\vx)$ is sufficiently smooth. Mild growth conditions on $\mD(\vx)$ and $\vb(\vx)$ as $|\vx|\to\infty$ can be imposed to guarantee that Eq. \eqref{fp-ic-eqn} is well-posed and admits a classical solution $f_{\alpha}$ in $\mathbb{R}^{N}\times(0,\infty)$. Moreover, the solution satisfies $f_{\alpha}>0$ in $\mathbb{R}^{N}\times(0,\infty)$, $\int_{\mathbb{R}^{N}}f_{\alpha}(\vx,t)\rd\vx=1$ for all $t>0$, and, together with its derivatives, decays to $0$ sufficiently fast as $|\vx|\to\infty$ \cite{jqq,Pavliotis2014}. 

\medskip

In addition, if $\vx_{\alpha}(t)$ is the solution to the SDE associated with Eq. \eqref{fp-eqn} with initial condition $\vx_{\alpha}(0)=\vx_0$, then for any $t>0$, $f_{\alpha}(\vx,t)$ is the density of the distribution of $\vx_{\alpha}(t)$.

\medskip

Under dissipative conditions (e.g., Lyapunov conditions), Eq. \eqref{fp-eqn} admits a unique positive stationary solution $\pi_{\alpha}$ satisfying $\int_{\mathbb{R}^{N}}\pi_{\alpha}\rd\vx=1$ \cite{Khasminskii12}. Setting 
\begin{equation}\label{stationary-prob-flux-append}
    \mJ_{\alpha}=\tfrac{1}{\alpha}\mD\nabla\ln\pi_{\alpha}-\vb, 
\end{equation}
then $\pi_{\alpha}$ satisfies
\begin{equation}\label{eqn-for-stationary-distribution-append}
    \nabla\cdot\left(\pi_{\alpha}\mJ_{\alpha}\right)=0.
\end{equation}

\medskip

Below, we state the Wentzel–Kramers–Brillouin (WKB) ansatz of $f_{\alpha}$ and $\pi_{\alpha}$ for large $\alpha$.

\subsection{WKB ansatz of $f_{\alpha}$}\label{appendix-WKB-transition-prob}

For each $t>0$, the WKB ansatz of $f_{\alpha}(\vx,t)$ in $\vx$ for large $\alpha$ reads (see e.g. \cite{Sheu84,fw,Friedman06})
\begin{equation}\label{WKB-f-alpha-append}
    f_{\alpha}=\left(\frac{\alpha}{2\pi}\right)^{N/2}R_{\alpha}e^{-\alpha\varphi}=\left(\frac{\alpha}{2\pi}\right)^{N/2}\left[R_{0}+\tfrac{1}{\alpha}R_{1}+O(\tfrac{1}{\alpha^2})\right]e^{-\alpha\varphi},
\end{equation}
where $\varphi=\varphi(\vx,t)$ is the rate function and $R_{\alpha}=R_{\alpha}(\vx,t)$ is the prefactor. The regularity of $\varphi$, $R_{\alpha}$, $R_{0}=R_0(\vx,t)$, and $R_1=R_{1}(\vx,t)$ is not automatic, but can be achieved under technical assumptions (see \cite[Section 3]{Sheu84}).

\medskip

The rate function $\varphi$ solves the Hamilton-Jacobi equation
$$
-\partial_{t}\varphi= \nabla\varphi\cdot\left(\mD\nabla\varphi + \vb\right).
$$
Moreover, if $\hat{\vx}(t)$ denotes the solution to the ODE $\dot{\vx}=\vb(\vx)$ with initial condition $\hat{\vx}(0)=\vx_{0}$, then for any $t>0$, the following hold:
\begin{itemize}
    \item $\varphi(\hat{\vx}(t),t)=0<\varphi(\vx,t)$ for all $\vx\in\mathbb{R}^{N}\setminus\{\hat{\vx}(t)\}$; in particular, $\nabla\varphi(\hat{\vx}(t),t)=0$;

    \item $\nabla\nabla\varphi(\hat{\vx}(t),t)$ is positive definite, where $\nabla\nabla\varphi$ denotes the Hessian matrix of $\varphi$.
\end{itemize}

\medskip

The leading term $R_{0}$ of the prefactor $R_{\alpha}$ solves the following linear equation
$$
-\partial_{t} R_{0}=\left(2\mD\nabla\varphi+\vb\right)\cdot\nabla R_{0}+\nabla\cdot\left(\mD\nabla\varphi+\vb\right)R_{0},
$$
while $R_1$ solves the following non-homogeneous equation
\begin{equation}\label{eqn-R1-append}
    -\partial_{t}R_1=-\nabla\cdot(\mD\nabla R_0)+\left(2\mD\nabla\varphi+\vb\right)\cdot\nabla R_{1}+\nabla\cdot\left(\mD\nabla\varphi+\vb\right)R_{1}.
\end{equation}


\medskip

Since $\left(\frac{\alpha}{2\pi}\right)^{N/2}\int_{\mathbb{R}^{N}}R_{\alpha}e^{-\alpha\varphi}\rd\vx=\int_{\mathbb{R}^{N}}f_{\alpha}\rd\vx=1$, Laplace's method yields
$$
\frac{R_{0}(\hat{\vx}(t),t)}{\sqrt{\det\nabla\nabla\varphi(\hat{\vx}(t),t)}}=1.
$$

\subsection{WKB ansatz of $\pi_{\alpha}$} 

The WKB ansatz of $\pi_{\alpha}$ reads
\begin{equation}\label{wkb-stationary-append}
\pi_{\alpha}=\frac{R_{\alpha}^{ss}}{C_{\alpha}}e^{-\alpha\varphi^{ss}},    
\end{equation}
where $C_{\alpha}$ is the sub-exponential (i.e., $\lim_{\alpha\to\infty}\tfrac{1}{\alpha}\ln C_{\alpha}=0$) normalizing constant, the rate function $\varphi^{ss}$ satisfies $\min\varphi^{ss}=0$ and solves the stationary Hamilton-Jacobi equation
\begin{equation}\label{eqn-rate-fun-ss}
    \nabla\varphi^{ss}\cdot\left(\mD\nabla\varphi^{ss}+\vb\right)=0,
\end{equation}
and the prefactor $R_{\alpha}^{ss}$ satisfies
    \begin{equation*}
R_{\alpha}^{ss}=R_{0}^{ss}+O\left(\tfrac{1}{\alpha}\right),
    \end{equation*}
in which $R_{0}^{ss}$ solves
$$
\left(2\mD\nabla\varphi^{ss}+\vb\right)\cdot\nabla R_{0}^{ss}+\left[\nabla\cdot\left(\mD\nabla\varphi^{ss}\right)+\nabla\cdot \vb\right]R_{0}^{ss}=0.
$$

\medskip

When $\vb$ admits a potential $U$, namely, $\vb=-\mD\nabla U$, then $\varphi=U$ and $R_{\alpha}^{ss}\equiv1$. This is the only trivial case. In general, the WKB ansatz of $\pi_{\alpha}$ relies heavily on the dynamical structure of the ODE 
\begin{equation}\label{ODE}
    \dot\vx=\vb(\vx).
\end{equation} 
The existence of the rate function $\varphi^{ss}$ has been justified under different dynamical assumptions on \eqref{ODE} by examining the limit of $-\tfrac{1}{\alpha}\ln\pi_{\alpha}$ as $\alpha\to\infty$. The asymptotic properties of $R_{\alpha}^{ss}$ are only known when \eqref{ODE} admit a globally asymptotically stable and non-degenerate equilibrium. See \cite{Sheu1986,Day1987,Mikami1990,BB2009}.

\section{Entropy}\label{entropy-append}

We present the entropy balance equation and study related large $\alpha$ asymptotics.

\subsection{Entropy balance equation}

The entropy $S=S[f_{\alpha}]$, entropy production rate $e_{p}=e_{p}[f_{\alpha}]$, and heat exchange rate $Q_{ex}=Q_{ex}[f_{\alpha}]$ are defined by
\begin{equation*}
    \begin{split}
        S&=-\int_{\mathbb{R}^{N}}f_{\alpha}\ln f_{\alpha}\rd\vx,\\
        e_p&= \int_{\mathbb{R}^N} 
       \left(\tfrac{1}{\alpha}\mD\nabla f_{\alpha}-\vb f_{\alpha}\right)\cdot\left(
          \nabla\ln f_{\alpha}- \alpha\mD^{-1}\vb \right)\rd\vx,\quad\text{and}\\
          Q_{ex}& =  \alpha\int_{\mathbb{R}^N} 
       \left(\tfrac{1}{\alpha}\mD\nabla f_{\alpha}-\vb f_{\alpha}\right)\cdot
           \left(\mD^{-1}\vb\right)\rd\vx,
    \end{split}
\end{equation*}
respectively. Whenever $f_{\alpha}$ is fixed, they are just functions of the time variable $t$.

\medskip

Clearly, $e_{p}$ can be written as
  \begin{equation}\label{entropy-production-rate-append}
      \begin{split}
          e_p 
          &=\alpha\int_{\mathbb{R}^N} 
       \left(\tfrac{1}{\alpha}\mD\nabla\ln f_{\alpha}-\vb\right)\cdot \mD^{-1}\left(
          \tfrac{1}{\alpha}\mD\nabla\ln f_{\alpha}- \vb \right)f_{\alpha}\rd\vx.
      \end{split}
  \end{equation}
The positive definiteness of $\mD^{-1}$ ensures that $e_{p}>0$. 

\medskip

Note that
  \begin{equation*}
      \begin{split}
          \frac{\rd S}{\rd t}&=-\int_{\mathbb{R}^{N}}\partial_{t} f_{\alpha}\left(\ln f_{\alpha}+1\right)\rd\vx\\
          &=-\int_{\mathbb{R}^{N}}\nabla\cdot \left( \tfrac{1}{\alpha}\mD\nabla f_{\alpha}-\vb f_{\alpha}\right)\left(\ln f_{\alpha}+1\right)\rd\vx\\
          &=\int_{\mathbb{R}^{N}}\left( \tfrac{1}{\alpha}\mD\nabla f_{\alpha}-\vb f_{\alpha}\right)\cdot\nabla\ln f_{\alpha}\rd\vx\\
          &=\int_{\mathbb{R}^{N}}\left( \tfrac{1}{\alpha}\mD\nabla f_{\alpha}-\vb f_{\alpha}\right)\cdot\left(\nabla\ln f_{\alpha}-\alpha\mD^{-1}\vb+\alpha\mD^{-1}\vb\right)\rd\vx.
      \end{split}
  \end{equation*}  
Hence, the following entropy balance equation holds:
\begin{equation}\label{entropy-balance-eqn-append}
\frac{\rd S}{\rd t}=e_{p}+Q_{ex}.    
\end{equation}

\subsection{Asymptotics} \label{app-B-2}

The large $\alpha$ asymptotics of $S$, $Q_{ex}$, $e_{p}$, and $\frac{dS}{dt}$ are given as follows.
\begin{equation*}
    \begin{split}
        S&=-\frac{N}{2}\ln\alpha+O(1),\\
        Q_{ex}&=-\alpha \vb(\hat{\vx}(t))\cdot \mD^{-1}(\hat{\vx}(t))\vb(\hat{\vx}(t))\\
        &\quad- C(\hat{\vx}(t),t)-\frac{R_{1}(\hat{\vx}(t),t)}{\sqrt{\det\nabla\nabla\varphi(\hat{\vx}(t),t)}}\vb(\hat{\vx}(t))\cdot \mD^{-1}(\hat{\vx}(t))\vb(\hat{\vx}(t))-\nabla\cdot\vb(\hat{\vx}(t))+O(\tfrac{1}{\alpha}),\\
        e_{p}&=\alpha \vb(\hat{\vx}(t))\cdot \mD^{-1}(\hat{\vx}(t))\vb(\hat{\vx}(t))\\
        &\quad+ C(\hat{\vx}(t),t)+\frac{R_{1}(\hat{\vx}(t),t)}{\sqrt{\det\nabla\nabla\varphi(\hat{\vx}(t),t)}}\vb(\hat{\vx}(t))\cdot \mD^{-1}(\hat{\vx}(t))\vb(\hat{\vx}(t))+2\nabla\cdot\vb(\hat{\vx}(t))\\
        &\quad+ \mD(\hat{\vx}(t)):\nabla\nabla\varphi(\hat{\vx}(t),t)+O(\tfrac{1}{\alpha}),\\
        \frac{\rd S}{\rd t}&=\mD(\hat{\vx}(t)):\nabla\nabla\varphi(\hat{\vx}(t),t)+\nabla\cdot\vb(\hat{\vx}(t))+O\left(\tfrac{1}{\alpha}\right),
    \end{split}
\end{equation*}
where the colon $:$ denotes the Frobenius product of two matrices of the same size, and
\begin{equation}\label{a-constant-append}
    C(\hat{\vx}(t),t)=\left[\frac{1}{2}\nabla\nabla(R_0\vb\cdot\mD^{-1}\vb):\nabla\nabla\varphi-\frac{1}{6}\nabla(R_0\vb\cdot\mD^{-1}\vb)\cdot(\nabla\nabla\nabla\varphi\Theta)\right]\bigg|_{\vx=\hat{\vx}(t)}.
\end{equation}
In which, $\nabla\nabla\nabla\varphi\Theta$ is an $N$-dimensional vector and its $i$-th component is given by
$$
(\nabla\nabla\nabla\varphi\Theta)_{i}=\sum_{j,k,\ell=1}^{N}\partial_{x_{j}x_{k}x_{\ell}}^{3}\varphi\Theta_{ijk\ell},
$$
where 
\begin{equation*}
       \Theta_{ijk\ell}
	 =  \frac{1}{(2\pi)^{N/2}\sqrt{\det(\nabla\nabla\varphi)^{-1}}}\int_{\mathbb{R}^N}
           y_iy_jy_ky_{\ell} e^{-\frac{\vy\cdot\nabla\nabla\varphi\vy}{2}}\rd\vy.
\end{equation*}

\medskip

Below, we provide the detailed derivation of these asymptotics.

\paragraph{\bf Asymptotic of $S$.} Clearly,
\begin{equation*}
    \begin{split}
        S&=-\left(\frac{\alpha}{2\pi}\right)^{N/2}\int_{\mathbb{R}^{N}}\left(\frac{N}{2}\ln\frac{\alpha}{2\pi}+\ln R_{\alpha}-\alpha\varphi\right)R_{\alpha}e^{-\alpha\varphi}\rd\vx\\
        &=-\frac{N}{2}\ln\frac{\alpha}{2\pi}-\ln R_{0}(\hat{\vx}(t),t)+O\left(\tfrac{1}{\alpha}\right)+\alpha\left(\frac{\alpha}{2\pi}\right)^{N/2}\int_{\mathbb{R}^{N}}\varphi R_{\alpha} e^{-\alpha\varphi}\rd\vx
    \end{split}
\end{equation*}
The asymptotic of $S$ follows from the following:
\begin{equation*}
    \begin{split}
        \alpha\left(\frac{\alpha}{2\pi}\right)^{N/2}\int_{\mathbb{R}^{N}}\varphi R_{\alpha} e^{-\alpha\varphi}\rd\vx&=-\left(\frac{\alpha}{2\pi}\right)^{N/2}\int_{\mathbb{R}^{N}}\frac{\varphi\nabla\varphi}{|\nabla\varphi|^2} R_{\alpha}\cdot \nabla e^{-\alpha\varphi}\rd\vx\\
        &=\left(\frac{\alpha}{2\pi}\right)^{N/2}\int_{\mathbb{R}^{N}}\nabla\cdot\left(\frac{\varphi\nabla\varphi}{|\nabla\varphi|^2} R_{\alpha}\right) e^{-\alpha\varphi}\rd\vx=O(1).
    \end{split}
\end{equation*}

\paragraph{\bf Asymptotic of $Q_{ex}$.} Since
\begin{equation*}
    \begin{split}
        \left(\tfrac{1}{\alpha}\mD\nabla f_{\alpha}-\vb f_{\alpha}\right)\cdot
           \left(\mD^{-1}\vb\right)&=\left(\frac{\alpha}{2\pi}\right)^{N/2}\left[\tfrac{1}{\alpha}\vb\cdot\nabla R_{\alpha}-\left(\vb\cdot\nabla\varphi+\vb\cdot \mD^{-1}\vb\right)R_{\alpha}\right]e^{-\alpha\varphi},
    \end{split}
\end{equation*}
we find
\begin{equation*}
    \begin{split}
        Q_{ex}
    &=-\alpha\left(\frac{\alpha}{2\pi}\right)^{N/2}\int_{\mathbb{R}^{N}}R_{\alpha}\vb\cdot\nabla\varphi e^{-\alpha\varphi}\rd\vx\\
    &\quad-\alpha\left(\frac{\alpha}{2\pi}\right)^{N/2}\int_{\mathbb{R}^{N}}R_{\alpha}\vb\cdot \mD^{-1}\vb e^{-\alpha\varphi}\rd\vx\\
    &\quad+\left(\frac{\alpha}{2\pi}\right)^{N/2}\int_{\mathbb{R}^{N}}\vb\cdot\nabla R_{\alpha}e^{-\alpha\varphi}\rd\vx\\
    &=I_{1}(\alpha)+I_{2}(\alpha)+I_{3}(\alpha).
    \end{split}
\end{equation*}

Note that 
\begin{equation*}
    \begin{split}
        I_{1}(\alpha)=\left(\frac{\alpha}{2\pi}\right)^{N/2}\int_{\mathbb{R}^{N}}R_{\alpha}\vb\cdot\nabla e^{-\alpha\varphi}\rd\vx=-\left(\frac{\alpha}{2\pi}\right)^{N/2}\int_{\mathbb{R}^{N}}\left(\nabla R_{\alpha}\cdot\vb+R_{\alpha}\nabla\cdot\vb\right) e^{-\alpha\varphi}\rd\vx.
    \end{split}
\end{equation*}
Then,
\begin{equation*}
    \begin{split}
        I_{1}(\alpha)+I_{3}(\alpha)=-\left(\frac{\alpha}{2\pi}\right)^{N/2}\int_{\mathbb{R}^{N}}R_{\alpha}\nabla\cdot\vb e^{-\alpha\varphi}\rd\vx=-\nabla\cdot\vb(\hat{\vx}(t))+O\left(\tfrac{1}{\alpha}\right)
    \end{split}
\end{equation*}

\medskip

For $I_{2}(\alpha)$, we see that
\begin{equation*}
    \begin{split}
        -\tfrac{1}{\alpha}I_{2}(\alpha)&=\left(\frac{\alpha}{2\pi}\right)^{N/2}\int_{\mathbb{R}^{N}}\left(R_{0}+\tfrac{1}{\alpha}R_{1}+O(\tfrac{1}{\alpha^2})\right)\vb\cdot \mD^{-1}\vb e^{-\alpha\varphi}\rd\vx\\
        &=\left(\frac{\alpha}{2\pi}\right)^{N/2}\int_{\mathbb{R}^{N}}R_{0}\vb\cdot \mD^{-1}\vb e^{-\alpha\varphi}\rd\vx\\
        &\quad+\tfrac{1}{\alpha}\left(\frac{\alpha}{2\pi}\right)^{N/2}\int_{\mathbb{R}^{N}}R_{1}\vb\cdot \mD^{-1}\vb e^{-\alpha\varphi}\rd\vx+O(\tfrac{1}{\alpha^2})\\
        &=I_{21}(\alpha)+\tfrac{1}{\alpha}I_{22}(\alpha)+O(\tfrac{1}{\alpha^2}).
    \end{split}
\end{equation*}
Since
\begin{equation*}
    \begin{split}
        I_{21}(\alpha)&=\vb(\hat{\vx}(t))\cdot \mD^{-1}(\hat{\vx}(t))\vb(\hat{\vx}(t))+\tfrac{1}{\alpha}C(\hat{\vx}(t),t)+O(\tfrac{1}{\alpha^2}),\\
        I_{22}(\alpha)&=\frac{R_{1}(\hat{\vx}(t),t)}{\sqrt{\det\nabla\nabla\varphi(\hat{\vx}(t),t)}}\vb(\hat{\vx}(t))\cdot \mD^{-1}(\hat{\vx}(t))\vb(\hat{\vx}(t))+O(\tfrac{1}{\alpha}),
    \end{split}
\end{equation*}
where $C(\hat{\vx}(t),t)$ is given in \eqref{a-constant-append}, we arrive at
\begin{equation*}
    \begin{split}
        I_{2}(\alpha)&=-\alpha \vb(\hat{\vx}(t))\cdot \mD^{-1}(\hat{\vx}(t))\vb(\hat{\vx}(t))\\
        &\quad- C(\hat{\vx}(t),t)-\frac{R_{1}(\hat{\vx}(t),t)}{\sqrt{\det\nabla\nabla\varphi(\hat{\vx}(t),t)}}\vb(\hat{\vx}(t))\cdot \mD^{-1}(\hat{\vx}(t))\vb(\hat{\vx}(t))+O(\tfrac{1}{\alpha}).
    \end{split}
\end{equation*}
The asymptotic of $Q_{ex}$ follows.

\paragraph{\bf Asymptotic of $e_{p}$.} Note that 
$$
e_{p}=-Q_{ex}+\underbrace{\int_{\mathbb{R}^{N}}\left( \tfrac{1}{\alpha}\mD\nabla f_{\alpha}-\vb f_{\alpha}\right)\cdot\nabla\ln f_{\alpha}\rd\vx}_{I(\alpha)}.
$$
Given the asymptotic of $Q_{ex}$, we only need to treat $I(\alpha)$. Straightforward calculations yield
\begin{equation*}
    \begin{split}
        &\left( \tfrac{1}{\alpha}\mD\nabla f_{\alpha}-\vb f_{\alpha}\right)\cdot\nabla\ln f_{\alpha}\\
        &\qquad=\left(\frac{\alpha}{2\pi}\right)^{N/2}\left[\tfrac{1}{\alpha}\frac{\nabla R_{\alpha}\cdot\mD\nabla R_{\alpha}}{R_{\alpha}}-\left(2\mD\nabla\varphi+\vb\right)\cdot\nabla R_{\alpha}+\alpha \left(\mD\nabla\varphi+\vb\right)R_{\alpha}\cdot\nabla\varphi\right]e^{-\alpha\varphi}.
    \end{split}
\end{equation*}
Then,
\begin{equation*}
    \begin{split}
        I(\alpha)&=\tfrac{1}{\alpha}\left(\frac{\alpha}{2\pi}\right)^{N/2}\int_{\mathbb{R}^{N}}\frac{\nabla R_{\alpha}\cdot\mD\nabla R_{\alpha}}{R_{\alpha}}e^{-\alpha\varphi}\rd\vx-\left(\frac{\alpha}{2\pi}\right)^{N/2}\int_{\mathbb{R}^{N}}\left(2\mD\nabla\varphi+\vb\right)\cdot\nabla R_{\alpha}e^{-\alpha\varphi}\rd\vx\\
        &\quad+\alpha\left(\frac{\alpha}{2\pi}\right)^{N/2}\int_{\mathbb{R}^{N}}R_{\alpha}\nabla\varphi\cdot\mD\nabla\varphi  e^{-\alpha\varphi}\rd\vx+\alpha\left(\frac{\alpha}{2\pi}\right)^{N/2}\int_{\mathbb{R}^{N}}R_{\alpha}\vb\cdot\nabla\varphi e^{-\alpha\varphi}\rd\vx.
    \end{split}
\end{equation*}
Note that the fourth term on the RHS of the above equality can be written as
\begin{equation*}
    \begin{split}
-\left(\frac{\alpha}{2\pi}\right)^{N/2}\int_{\mathbb{R}^{N}}R_{\alpha}\vb\cdot\nabla e^{-\alpha\varphi}\rd\vx=\left(\frac{\alpha}{2\pi}\right)^{N/2}\int_{\mathbb{R}^{N}}\left(\nabla R_{\alpha}\cdot\vb+R_{\alpha}\nabla\cdot\vb\right) e^{-\alpha\varphi}\rd\vx.
    \end{split}
\end{equation*}
Hence,
\begin{equation*}
    \begin{split}
        I(\alpha)&=\tfrac{1}{\alpha}\left(\frac{\alpha}{2\pi}\right)^{N/2}\int_{\mathbb{R}^{N}}\frac{\nabla R_{\alpha}\cdot\mD\nabla R_{\alpha}}{R_{\alpha}}e^{-\alpha\varphi}\rd\vx-2\left(\frac{\alpha}{2\pi}\right)^{N/2}\int_{\mathbb{R}^{N}}\mD\nabla\varphi\cdot\nabla R_{\alpha}e^{-\alpha\varphi}\rd\vx\\
        &\quad+\alpha\left(\frac{\alpha}{2\pi}\right)^{N/2}\int_{\mathbb{R}^{N}}R_{\alpha}\nabla\varphi\cdot\mD\nabla\varphi  e^{-\alpha\varphi}\rd\vx+\left(\frac{\alpha}{2\pi}\right)^{N/2}\int_{\mathbb{R}^{N}}R_{\alpha}\nabla\cdot\vb e^{-\alpha\varphi}\rd\vx\\
        &=I_{4}(\alpha)+I_{5}(\alpha)+I_{6}(\alpha)+I_{7}(\alpha).
    \end{split}
\end{equation*}
Clearly, $I_{4}(\alpha)=O\left(\tfrac{1}{\alpha}\right)$ and $I_{7}(\alpha)=\nabla\cdot\vb(\hat{\vx}(t))+O\left(\tfrac{1}{\alpha}\right)$. For $I_{5}(\alpha)$,
\begin{equation*}
    \begin{split}
        I_{5}(\alpha)&=\tfrac{2}{\alpha}\left(\frac{\alpha}{2\pi}\right)^{N/2}\int_{\mathbb{R}^{N}}\mD\nabla R_{\alpha}\cdot\nabla e^{-\alpha\varphi}\rd\vx\\
        &=-\tfrac{2}{\alpha}\left(\frac{\alpha}{2\pi}\right)^{N/2}\int_{\mathbb{R}^{N}}\nabla\cdot\left(\mD\nabla R_{\alpha}\right) e^{-\alpha\varphi}\rd\vx=O\left(\tfrac{1}{\alpha}\right).
    \end{split}
\end{equation*}

For $I_{6}(\alpha)$, 
\begin{equation*}
    \begin{split}
        I_{6}(\alpha)&=-\left(\frac{\alpha}{2\pi}\right)^{N/2}\int_{\mathbb{R}^{N}}R_{\alpha}\mD\nabla\varphi\cdot \nabla e^{-\alpha\varphi}\rd\vx\\
        &=\left(\frac{\alpha}{2\pi}\right)^{N/2}\int_{\mathbb{R}^{N}}\nabla\cdot\left(R_{\alpha}\mD\nabla\varphi\right) e^{-\alpha\varphi}\rd\vx\\
        &=\left(\frac{\alpha}{2\pi}\right)^{N/2}\int_{\mathbb{R}^{N}}\left[\nabla R_{\alpha}\cdot\mD\nabla\varphi+R_{\alpha}\left(\nabla\cdot\mD\right)\cdot\nabla\varphi+R_{\alpha}\mD:\nabla\nabla\varphi\right] e^{-\alpha\varphi}\rd\vx\\
        &=-\tfrac{1}{\alpha}\left(\frac{\alpha}{2\pi}\right)^{N/2}\int_{\mathbb{R}^{N}}\left[\mD\nabla R_{\alpha}+R_{\alpha}\left(\nabla\cdot\mD\right)\right]\cdot\nabla e^{-\alpha\varphi}\rd\vx+\left(\frac{\alpha}{2\pi}\right)^{N/2}\int_{\mathbb{R}^{N}}R_{\alpha}\mD:\nabla\nabla\varphi e^{-\alpha\varphi}\rd\vx\\
        &=\tfrac{1}{\alpha}\left(\frac{\alpha}{2\pi}\right)^{N/2}\int_{\mathbb{R}^{N}}\nabla\cdot\left[\mD\nabla R_{\alpha}+R_{\alpha}\left(\nabla\cdot\mD\right)\right] e^{-\alpha\varphi}\rd\vx+\left(\frac{\alpha}{2\pi}\right)^{N/2}\int_{\mathbb{R}^{N}}R_{\alpha}\mD:\nabla\nabla\varphi e^{-\alpha\varphi}\rd\vx\\
        &=O\left(\tfrac{1}{\alpha}\right)+\mD(\hat{\vx}(t)):\nabla\nabla\varphi(\hat{\vx}(t),t).
    \end{split}
\end{equation*}
Hence,
\begin{equation*}
    \begin{split}
I(\alpha)=\mD(\hat{\vx}(t)):\nabla\nabla\varphi(\hat{\vx}(t),t)+\nabla\cdot\vb(\hat{\vx}(t))+O\left(\tfrac{1}{\alpha}\right).        
    \end{split}
\end{equation*}
The asymptotic of $e_{p}$ then follows.

\paragraph{\bf Asymptotic of $\frac{\rd S}{\rd t}$.} It follows from the entropy balance equation Eq. \eqref{entropy-balance-eqn-append} and the asymptotics of $e_{p}$ and $Q_{ex}$. Alternatively, the asymptotic of $\frac{\rd S}{\rd t}$ follows from the fact that $\frac{\rd S}{\rd t}=I(\alpha)$.


\section{Free energy}

We present the free energy balance equation and study related large $\alpha$ asymptotics.

\subsection{Free energy balance equation}\label{appendix-free-energy-balance-eqn}

The free energy $F=F[f_{\alpha}]$ and house-keeping heat rate $Q_{hk}=Q_{hk}[f_{\alpha}]$ are defined by
\begin{equation*}
    \begin{split}
        F&=\int_{\mathbb{R}^{N}}f_{\alpha}\ln \frac{f_{\alpha}}{\pi_{\alpha}}\rd\vx\quad\text{and}\\
        Q_{hk}&=\int_{\mathbb{R}^N} 
       \left(\tfrac{1}{\alpha}\mD\nabla f_{\alpha}-\vb f_{\alpha}\right)\cdot\left(
          \nabla\ln \pi_{\alpha}- \alpha\mD^{-1}\vb \right)\rd\vx,
    \end{split}
\end{equation*}
respectively. It is known that $F>0$ and $Q_{hk}\geq0$, and the following free energy balance equation holds:
\begin{equation}\label{free-energy-balance-eqn-append}
\frac{\rd F}{\rd t}=-e_{p}+Q_{hk}.    
\end{equation}

Below, we provide the details.


\paragraph{\bf Positivity of $F$.} Note that 
$$
F=-\int_{\mathbb{R}^{N}}f_{\alpha}\ln \frac{\pi_{\alpha}}{f_{\alpha}}\rd\vx\geq-\ln\int_{\mathbb{R}^{N}}\frac{\pi_{\alpha}}{f_{\alpha}}f_{\alpha}\rd\vx=0,
$$
where we used Jensen's inequality. Moreover, since the function $x\mapsto -\ln x$ is strictly convex, the above inequality is strict unless $f_{\alpha}\equiv\pi_{\alpha}$. As $f_{\alpha}\neq\pi_{\alpha}$, one concludes $F>0$.


\paragraph{\bf Non-negativity of $Q_{hk}$.} We show
\begin{equation}\label{house-keep-alternative-append}
   Q_{hk}=\alpha\int_{\mathbb{R}^N}\mJ_{\alpha} \cdot\mD^{-1}\mJ_{\alpha}f_{\alpha}\rd\vx\geq0,    
\end{equation}
where $\mJ_{\alpha}$ is defined in \eqref{stationary-prob-flux-append}. Note that $\mJ_{\alpha}=\bf 0$ so that $Q_{hk}=0$ when $\vb$ admits a potential $U$, namely, $\vb=-\mD\nabla U$. Otherwise, $Q_{hk}>0$.

\medskip

Indeed, we calculate
\begin{equation*}
    \begin{split}
        Q_{hk}
       &=\alpha\int_{\mathbb{R}^N} 
       \left(\tfrac{1}{\alpha}\mD\nabla \ln f_{\alpha}-\vb \right)f_{\alpha}\cdot\mD^{-1}\mJ_{\alpha}\rd\vx\\
       &=\alpha\int_{\mathbb{R}^N} 
       \left(\tfrac{1}{\alpha}\mD\nabla \ln f_{\alpha}-\tfrac{1}{\alpha}\mD\nabla\ln \pi_{\alpha}+\mJ_{\alpha} \right)f_{\alpha}\cdot\mD^{-1}\mJ_{\alpha}\rd\vx\\
       &=\alpha\int_{\mathbb{R}^N} 
       \left(\tfrac{1}{\alpha}\mD\nabla \ln f_{\alpha}-\tfrac{1}{\alpha}\mD\nabla\ln \pi_{\alpha} \right)f_{\alpha}\cdot\mD^{-1}\mJ_{\alpha}\rd\vx+\alpha\int_{\mathbb{R}^N} 
       \mJ_{\alpha} \cdot\mD^{-1}\mJ_{\alpha}f_{\alpha}\rd\vx\\
       &=\int_{\mathbb{R}^N}\mD\nabla \ln \frac{f_{\alpha}}{\pi_{\alpha}}f_{\alpha}\cdot\mD^{-1}\mJ_{\alpha}\rd\vx+\alpha\int_{\mathbb{R}^N} 
       \mJ_{\alpha} \cdot\mD^{-1}\mJ_{\alpha}f_{\alpha}\rd\vx.
    \end{split}
\end{equation*}
The expected result follows readily from
\begin{equation}\label{identity-2024-03-21}
\int_{\mathbb{R}^N}\mD\nabla \ln \frac{f_{\alpha}}{\pi_{\alpha}}f_{\alpha}\cdot\mD^{-1}\mJ_{\alpha}\rd\vx=\int_{\mathbb{R}^N}\nabla \frac{f_{\alpha}}{\pi_{\alpha}}\cdot\left(\pi_{\alpha}\mJ_{\alpha}\right)\rd\vx=-\int_{\mathbb{R}^N}\frac{f_{\alpha}}{\pi_{\alpha}}\nabla\cdot\left(\pi_{\alpha}\mJ_{\alpha}\right)\rd\vx=0,   
\end{equation}
where we used Eq. \eqref{eqn-for-stationary-distribution-append} in the last equality.


\paragraph{\bf Free energy balance equation.}
We show
\begin{equation*}
    \begin{split}
        \frac{\rd F}{\rd t}=-e_{p}+Q_{hk}=-\tfrac{1}{\alpha}\int_{\mathbb{R}^N} 
       \left(\nabla\ln \frac{f_{\alpha}}{\pi_{\alpha}}\right)\cdot \mD\left(\nabla\ln\frac{f_{\alpha}}{\pi_{\alpha}} \right)f_{\alpha}\rd\vx<0.
    \end{split}
\end{equation*}
The strict inequality is a result of the fact that $f_{\alpha}\neq\pi_{\alpha}$.

\medskip

The first equality follows readily.
 \begin{equation*}
      \begin{split}
          \frac{\rd F}{\rd t}&=\int_{\mathbb{R}^{N}}\partial_{t} f_{\alpha}\left(\ln \frac{f_{\alpha}}{\pi_{\alpha}}+1\right)\rd\vx\\
          &=\int_{\mathbb{R}^{N}}\nabla\cdot \left( \tfrac{1}{\alpha}\mD\nabla f_{\alpha}-\vb f_{\alpha}\right)\left(\ln \frac{f_{\alpha}}{\pi_{\alpha}}+1\right)\rd\vx\\
          &=-\int_{\mathbb{R}^{N}}\left( \tfrac{1}{\alpha}\mD\nabla f_{\alpha}-\vb f_{\alpha}\right)\cdot\left(\nabla\ln f_{\alpha}-\nabla\ln\pi_{\alpha}\right)\rd\vx\\
          &=-\int_{\mathbb{R}^{N}}\left( \tfrac{1}{\alpha}\mD\nabla f_{\alpha}-\vb f_{\alpha}\right)\cdot\left(\nabla\ln f_{\alpha}-\alpha\mD^{-1}\vb+\alpha\mD^{-1}\vb-\nabla\ln\pi_{\alpha}\right)\rd\vx\\
          &=-e_{p}+Q_{hk}.
      \end{split}
  \end{equation*} 

\medskip

Now, we treat the second equality. From \eqref{entropy-production-rate-append}, one deduces
\begin{equation*}
      \begin{split}
          e_p 
          &=\alpha\int_{\mathbb{R}^N} 
       \left(\tfrac{1}{\alpha}\mD\nabla\ln \frac{f_{\alpha}}{\pi_{\alpha}}+\mJ_{\alpha}\right)\cdot \mD^{-1}\left(
          \tfrac{1}{\alpha}\mD\nabla\ln\frac{f_{\alpha}}{\pi_{\alpha}}+\mJ_{\alpha} \right)f_{\alpha}\rd\vx\\
          &=\tfrac{1}{\alpha}\int_{\mathbb{R}^N} 
       \left(\mD\nabla\ln \frac{f_{\alpha}}{\pi_{\alpha}}\right)\cdot \mD^{-1}\left(\mD\nabla\ln\frac{f_{\alpha}}{\pi_{\alpha}} \right)f_{\alpha}\rd\vx\\
       &\quad+2\int_{\mathbb{R}^N} 
       \left(\mD\nabla\ln \frac{f_{\alpha}}{\pi_{\alpha}}\right)\cdot \mD^{-1}\mJ_{\alpha}f_{\alpha}\rd\vx+\alpha\int_{\mathbb{R}^N} 
       \mJ_{\alpha}\cdot \mD^{-1}\mJ_{\alpha} f_{\alpha}\rd\vx\\
       &=\tfrac{1}{\alpha}\int_{\mathbb{R}^N} 
       \left(\nabla\ln \frac{f_{\alpha}}{\pi_{\alpha}}\right)\cdot \mD\left(\nabla\ln\frac{f_{\alpha}}{\pi_{\alpha}} \right)f_{\alpha}\rd\vx+Q_{hk},
      \end{split}
  \end{equation*}
where we used Eq. \eqref{house-keep-alternative-append} and Eq. \eqref{identity-2024-03-21} in the last equality. The free energy balance equation follows.

\medskip

Rewriting 
$$
\frac{\rd F}{\rd t}=-\alpha\int_{\mathbb{R}^N} 
       \left(\tfrac{1}{\alpha}\mD\nabla\ln \frac{f_{\alpha}}{\pi_{\alpha}}\right)\cdot \mD^{-1}\left(\tfrac{1}{\alpha}\mD\nabla\ln\frac{f_{\alpha}}{\pi_{\alpha}} \right)f_{\alpha}\rd\vx
$$ 
and introducing the inner product and the associated norm
$$
\left\langle\mV_1,\mV_2\right\rangle_{\smD^{-1},f_{\alpha}}=\int_{\mathbb{R}^{N}}\mV_1\cdot\mD^{-1}\mV_2 f_{\alpha}\rd\vx\quad\text{and}\quad \|\mV\|_{\smD^{-1},f_{\alpha}}=\sqrt{\left\langle\mV,\mV\right\rangle_{\smD^{-1},f_{\alpha}}}
$$
for vector fields $\mV_1$, $\mV_2$ and $\mV$ on $\mathbb{R}^{N}$, the free energy balance equation can be rewritten as
\begin{equation}\label{free-energy-balance-eqn-in-quadratic-form-append}
    \left\|\tfrac{1}{\alpha}\mD\nabla\ln \frac{f_{\alpha}}{\pi_{\alpha}}\right\|_{\smD^{-1},f_{\alpha}}^{2}+\left\|\mJ_{\alpha}\right\|_{\smD^{-1},f_{\alpha}}^{2}=\left\|\tfrac{1}{\alpha}\mD\nabla\ln \frac{f_{\alpha}}{\pi_{\alpha}}+\mJ_{\alpha}\right\|_{\smD^{-1},f_{\alpha}}^{2},
\end{equation}
which is equivalent to the orthogonality between $\mD\nabla\ln \frac{f_{\alpha}}{\pi_{\alpha}}$ and $\mJ_{\alpha}$ w.r.t. the inner product, that is,
$$
\left\langle\mD\nabla\ln \frac{f_{\alpha}}{\pi_{\alpha}},\mJ_{\alpha}\right\rangle_{\smD^{-1},f_{\alpha}}=0.
$$
This is just Eq. \eqref{identity-2024-03-21}.

\subsection{Asymptotics}\label{appendix-asymptotic-generalized-free-energy}

The asymptotics of $F$, $Q_{hk}$, and $\frac{\rd F}{\rd t}$ are given as follows:
\begin{equation*}
    \begin{split}
        F&=\alpha\varphi^{ss}(\hat{\vx}(t))+\frac{N}{2}\ln\alpha-\ln C_{\alpha}+O(1),\\
        Q_{hk}&=\alpha \vgamma(\hat{\vx}(t))\cdot \mD^{-1}(\hat{\vx}(t))\vgamma(\hat{\vx}(t))+O(1),\\
        \frac{\rd F}{\rd t}&=-\alpha\nabla\varphi^{ss}(\hat{\vx}(t))\cdot \mD(\hat{\vx}(t))\nabla\varphi^{ss}(\hat{\vx}(t))+O(1),
    \end{split}
\end{equation*}
where $C_{\alpha}$ is given in \eqref{wkb-stationary-append} and $\vgamma=\mD\nabla\varphi^{ss}+\vb$. If \eqref{ODE} admit a globally asymptotically stable and non-degenerate equilibrium, then $C_{\alpha}=\left(\tfrac{\alpha}{2\pi}\right)^{-N/2}$, resulting in $F=\alpha\varphi^{ss}(\hat{\vx}(t))+O(1)$.

\medskip

Multiplying the free energy balance equation Eq. \eqref{free-energy-balance-eqn-append} by $\tfrac{1}{\alpha}$ and then letting $\alpha\to\infty$, we derive from the asymptotics of $\frac{dF}{dt}$, $e_{p}$, and $Q_{hk}$ the following free energy balance equation on the macroscopic scale: 
$$
-\nabla\varphi^{ss}(\vx)\cdot \mD(\vx)\nabla\varphi^{ss}(\vx)=-\vb(\vx)\cdot \mD^{-1}(\vx)\vb(\vx)+\vgamma(\vx)\cdot \mD^{-1}(\vx)\vgamma(\vx) \quad\text{at}\quad \vx=\hat{\vx}(t).
$$
Introducing the norm $\|\vv\|_{\smD^{-1}(\vx)}=\sqrt{\vv\cdot\mD^{-1}(\vx)\vv}$ for $\vv\in\mathbb{R}^{N}$, we arrive at the Pythagorean equality 
$$
\|\mD(\vx)\nabla\varphi^{ss}(\vx)\|_{\smD^{-1}(\vx)}^{2}+\|\vgamma(\vx)\|_{\smD^{-1}(\vx)}^{2}=\|\vb(\vx)\|_{\smD^{-1}(\vx)}^{2}\quad\text{at}\quad \vx=\hat{\vx}(t),
$$
which is equivalent to Eq. \eqref{eqn-rate-fun-ss} (or $\nabla\varphi^{ss}\cdot\gamma=0$) at $\hat{\vx}(t)$. Of course, it is just the leading asymptotic of Eq. \eqref{free-energy-balance-eqn-in-quadratic-form-append}.

\medskip

Below, we justify the asymptotics of $F$, $Q_{hk}$, and $\frac{dF}{dt}$.

\paragraph{\bf Asymptotic of $F$.} Clearly,
\begin{equation*}
    \begin{split}
        F&=S-\int_{\mathbb{R}^{N}}f_{\alpha}\ln\pi_{\alpha}\rd\vx\\
        &=\frac{N}{2}\ln\alpha+O(1)-\left(\frac{\alpha}{2\pi}\right)^{N/2}\int_{\mathbb{R}^{N}}\left(\ln C_{\alpha}+\ln R_{\alpha}^{ss}-\alpha\varphi^{ss}\right)R_{\alpha}e^{-\alpha\varphi}\rd\vx\\
        &=\alpha\varphi^{ss}(\hat{\vx}(t))+\frac{N}{2}\ln\alpha-\ln C_{\alpha}+O(1).
    \end{split}
\end{equation*}

\paragraph{\bf Asymptotic of $Q_{hk}$.} 
Since
$$
\mJ_{\alpha}=\tfrac{1}{\alpha}\frac{\mD\nabla R_{\alpha}^{ss}}{R_{\alpha}^{ss}}-\mD\nabla\varphi^{ss}-\vb=\tfrac{1}{\alpha}\frac{\mD\nabla R_{\alpha}^{ss}}{R_{\alpha}^{ss}}-\vgamma,
$$
we find
\begin{equation*}
    \begin{split}
        Q_{hk}&=\alpha\left(\frac{\alpha}{2\pi}\right)^{N/2}\int_{\mathbb{R}^{N}}\left(\tfrac{1}{\alpha}\frac{\mD\nabla R_{\alpha}^{ss}}{R_{\alpha}^{ss}}-\vgamma\right)\cdot\mD^{-1}\left(\tfrac{1}{\alpha}\frac{\mD\nabla R_{\alpha}^{ss}}{R_{\alpha}^{ss}}-\vgamma\right)R_{\alpha}e^{-\alpha\varphi}\rd\vx\\
        &=\tfrac{1}{\alpha}\left(\frac{\alpha}{2\pi}\right)^{N/2}\int_{\mathbb{R}^{N}}\frac{\nabla R_{\alpha}^{ss}\cdot\mD\nabla R_{\alpha}^{ss}}{(R_{\alpha}^{ss})^{2}}R_{\alpha}e^{-\alpha\varphi}\rd\vx\\
        &\quad+2\left(\frac{\alpha}{2\pi}\right)^{N/2}\int_{\mathbb{R}^{N}}\frac{\nabla R_{\alpha}^{ss}\cdot\vgamma}{R_{\alpha}^{ss}}R_{\alpha}e^{-\alpha\varphi}\rd\vx\\
        &\quad+\alpha\left(\frac{\alpha}{2\pi}\right)^{N/2}\int_{\mathbb{R}^{N}}\gamma\cdot\mD^{-1}\vgamma R_{\alpha}e^{-\alpha\varphi}\rd\vx\\ 
        &=\alpha \vgamma(\hat{\vx}(t))\cdot \mD^{-1}(\hat{\vx}(t))\vgamma(\hat{\vx}(t))+O(1).
    \end{split}
\end{equation*}

\paragraph{\bf Asymptotic of $\frac{\rd F}{\rd t}$.}   
Since
    \begin{equation*}
        \begin{split}
            \nabla\ln \frac{f_{\alpha}}{\pi_{\alpha}}&=\frac{\nabla R_{\alpha}}{R_{\alpha}}-\frac{\nabla R_{\alpha}^{ss}}{R_{\alpha}^{ss}}-\alpha\left(\nabla\varphi-\nabla\varphi^{ss}\right),
        \end{split}
    \end{equation*}
we find
\begin{equation*}
    \begin{split}
        \frac{\rd F}{\rd t}&=-\tfrac{1}{\alpha}\int_{\mathbb{R}^N} 
       \left(\frac{\nabla R_{\alpha}}{R_{\alpha}}-\frac{\nabla R_{\alpha}^{ss}}{R_{\alpha}^{ss}}\right)\cdot \mD\left(\frac{\nabla R_{\alpha}}{R_{\alpha}}-\frac{\nabla R_{\alpha}^{ss}}{R_{\alpha}^{ss}} \right)f_{\alpha}\rd\vx\\
       &\quad+2\int_{\mathbb{R}^N} 
       \left(\nabla\varphi-\nabla\varphi^{ss}\right)\cdot \mD\left(\frac{\nabla R_{\alpha}}{R_{\alpha}}-\frac{\nabla R_{\alpha}^{ss}}{R_{\alpha}^{ss}} \right)f_{\alpha}\rd\vx\\
       &\quad-\alpha\int_{\mathbb{R}^N} 
       \left(\nabla\varphi-\nabla\varphi^{ss}\right)\cdot \mD\left(\nabla\varphi-\nabla\varphi^{ss}\right)f_{\alpha}\rd\vx\\
       &=O(1)-\alpha\int_{\mathbb{R}^N} 
       \nabla\varphi\cdot \mD\nabla\varphi f_{\alpha}\rd\vx+2\alpha\int_{\mathbb{R}^N} 
       \nabla\varphi\cdot \mD\nabla\varphi^{ss}f_{\alpha}\rd\vx-\alpha\int_{\mathbb{R}^N} 
       \nabla\varphi^{ss}\cdot \mD\nabla\varphi^{ss}f_{\alpha}\rd\vx\\
       &=O(1)+I_{8}(\alpha)+I_{9}(\alpha)+I_{10}(\alpha).
    \end{split}
\end{equation*}
Clearly,
$$
I_{10}(\alpha)=-\alpha\nabla\varphi^{ss}(\hat{\vx}(t))\cdot \mD(\hat{\vx}(t))\nabla\varphi^{ss}(\hat{\vx}(t))+O(1).
$$

For $I_{9}(\alpha)$,
\begin{equation*}
    \begin{split}
        I_{9}(\alpha)&=-2\left(\frac{\alpha}{2\pi}\right)^{N/2}\int_{\mathbb{R}^{N}}\mD\nabla\varphi^{ss}R_{\alpha}\cdot\nabla e^{-\alpha\varphi}\rd\vx\\
        &=2\left(\frac{\alpha}{2\pi}\right)^{N/2}\int_{\mathbb{R}^{N}}\nabla\cdot\left(\mD\nabla\varphi^{ss}R_{\alpha}\right)e^{-\alpha\varphi}\rd\vx=O(1).
    \end{split}
\end{equation*}

For $I_{8}(\alpha)$,
\begin{equation*}
    \begin{split}
        I_{8}(\alpha)&=\left(\frac{\alpha}{2\pi}\right)^{N/2}\int_{\mathbb{R}^{N}}\mD\nabla\varphi R_{\alpha}\cdot\nabla e^{-\alpha\varphi}\rd\vx\\
        &=-\left(\frac{\alpha}{2\pi}\right)^{N/2}\int_{\mathbb{R}^{N}}\nabla\cdot\left(\mD\nabla\varphi R_{\alpha}\right) e^{-\alpha\varphi}\rd\vx=O(1).
    \end{split}
\end{equation*}
Hence, the expected asymptotic of $\frac{\rd F}{\rd t}$ follows.


\section{Ornstein-Uhlenbeck process}
\label{app_OU}

Consider the Ornstein-Uhlenbeck process with linear drift $\vb(\vx)=\mB\vx$ and constant diffusion matrix $\tfrac{1}{\alpha}\mD$ (symmetric and positive definite), where $\mB$ is a $N\times N$ matrix. Then,
$$
f_{\alpha}(\vx,t)=\frac{1}{(2\pi)^{N/2}}\frac{1}{\sqrt{\det\mSigma_{\alpha}(t)}}\exp\left\{-\tfrac{1}{2}\left(\vx-e^{\mB t}\vx_0\right)\cdot\mSigma_{\alpha}^{-1}(t)\left(\vx-e^{\mB t}\vx_0\right)\right\},
$$
where the covariance matrix $\mSigma_{\alpha}(t)$ is given by
$$
\mSigma_{\alpha}(t)=\frac{2}{\alpha}\int_{0}^{t}e^{\mB(t-s)}\mD e^{\mB^{\top}(t-s)}ds.
$$
Introducing
$$
\mSigma(t)=2\int_{0}^{t}e^{\mB(t-s)}\mD e^{\mB^{\top}(t-s)}ds,
$$
one can rewrite $f_{\alpha}$ as
$$
f_{\alpha}(\vx,t)=\left(\frac{\alpha}{2\pi}\right)^{N/2}\frac{1}{\sqrt{\det\mSigma(t)}}\exp\left\{-\tfrac{\alpha}{2}\left(\vx-e^{\mB t}\vx_0\right)\cdot\mSigma^{-1}(t)\left(\vx-e^{\mB t}\vx_0\right)\right\}.
$$
The WKB ansatz \eqref{WKB-f-alpha-append} holds precisely in this case:
\begin{equation}\label{precise-WKB-ansatz-OU}
    f_{\alpha}(\vx,t)=\left(\frac{\alpha}{2\pi}\right)^{N/2}R_{0}(\vx,t)e^{-\alpha\varphi(\vx,t)},
\end{equation}
where $R_{0}(\vx,t)=\frac{1}{\sqrt{\det\mSigma(t)}}$ is independent of $\vx$ and 
$$
\varphi(\vx,t)=\tfrac{1}{2}\left(\vx-e^{\mB t}\vx_0\right)\cdot\mSigma^{-1}(t)\left(\vx-e^{\mB t}\vx_0\right).
$$

\medskip

It is well-known that 
\begin{equation*}
    \begin{split}
        S=\frac{N}{2}\ln\left(2\pi e\right)+\frac{1}{2}\ln\left(\det\mSigma_{\alpha}(t)\right)=\frac{N}{2}\ln\left(2\pi e\right)+\frac{1}{2}\ln\left(\det\mSigma(t)\right)-\frac{N}{2}\ln\alpha.
    \end{split}
\end{equation*}
It follows from Jacobi's formula and the formula $\mSigma'(t)=2\mD+\mB\mSigma(t)+\mSigma(t)\mB^{\top}$ \cite{qian-prsa-01} that
\begin{equation*}
    \begin{split}
\frac{\rd S}{\rd t}&=\frac{1}{2}\frac{\rd}{\rd t}\ln\left(\det\mSigma(t)\right)\\
&=\frac{1}{2}\text{tr}\left[\mSigma^{-1}(t)\mSigma'(t)\right]\\
&=\text{tr}\left[\mD\mSigma^{-1}(t)\right]+\text{tr}\mB\\
&=\mD:\nabla\nabla\varphi+\nabla\cdot\vb.
    \end{split}
\end{equation*}


\begin{thebibliography}{99}


\bibitem{alekseev-yakobson}
Alekseev, V. M. and Yakobson, M .V. (1981)
Symbolic dynamics and hyperbolic dynamical systems.
{\em Phys. Rep.} {\bf 75}, 287--325.

\bibitem{andrey}
Andrey, L. (1985)
The rate of entropy change in non-Hamiltonian systems.
{\em Phys. Lett. A} {\bf 111}, 45--46.

\bibitem{ben-naim}
Ben-Naim, A. (2006)
The entropy of mixing and assimilation: An 
information-theoretical perspective.
{\em Am. J. Phys.} {\bf 74}, 1126--1135.


\bibitem{BL55}
Bergmann, P. G.  and Lebowitz, J. L. (1955)
New Approach to Nonequilibrium Processes. {\em Phys. Rev.} {\bf 99}, 578--587.

\bibitem{BB2009}
Biswas, A. and Borkar, V. S. (2009) 
Small noise asymptotics for invariant densities for a class of diffusions: a control theoretic view. {\em J. Math. Anal. Appl.} {\bf 360}, 476--484. 

\bibitem{cqz-2020}
Cheng, Y.-C., Qian, H. and Zhu, Y. (2021)
Asymptotic behavior of a sequence of conditional probability distributions and the canonical ensemble.
{\em Ann. Henri Poincar\'e} {\bf 22}, 1561--1627.

\bibitem{cheng_qian_jsp}
Cheng, Y.-C. and Qian, H. (2021)
Stochastic limit-cycle oscillations of a nonlinear system under random perturbations.
{\em J. Stat. Phys.} {\bf 182}, 47.


\bibitem{cohen_rondoni}
Cohen, E. G. D. and Rondoni, L. (1998) Note on phase space contraction and entropy production in thermostatted Hamiltonian systems. {\em Chaos} {\bf 8}, 357--365.


\bibitem{CT05}
Cover, T. M. and Thomas, J. A. (2005) 
{\em Elements of Information Theory.} 2$^{nd}$ ed., John Wiley \& Sons, Inc.

\bibitem{young2}
Cowieson, W. and Young, L.-S. (2005) 
SRB measures as zero-noise limits.
{\em Erg. Th. Dyn. Sys.}  {\bf 25}, 1115--1138.

\bibitem{Day1987}
Day, M. V. (1987)
Recent progress on the small parameter exit problem. 
{\em Stochastics} {\bf 20}, 121--150.


\bibitem{degroot1962}
de Groot, S. R. and Mazur, P. (1962)
{\em Non-Equilibrium Thermodynamics},
North-Holland, Amsterdam. 

\bibitem{dorfman}
Dorfman, J. R. (1999)
{\em An Introduction to Chaos in Nonequilibrium Statistical
Mechanics}, Cambridge Univ. Press, London.

\bibitem{esposito-prl}
Esposito, M. and Van den Broeck, C. (2010)
Three detailed fluctuation theorems.
{\em Phys. Rev. Lett.} {\bf 104}, 090601.

\bibitem{FE24}
Falasco, G. and Esposito, M. (2025)
Macroscopic stochastic thermodynamics. \emph{Rev. Mod. Phys.} {\bf 97}, 015002.

\bibitem{FE22}
Freitas, J. N. and Esposito, M.  (2022)
Emergent second law for non-equilibrium steady states. {\em Nat Commun} {\bf 13}, 5084.

\bibitem{fw}
Freidlin, M. I. and Wentzell, A. D. (1998)
{\em Random Perturbations of Dynamical Systems},
2$^{nd}$ ed.,  Springer-Verlag, New York.

\bibitem{Friedman06}
Friedman, A. (2006) 
{\em Stochastic Differential Equations and Applications}. Dover, New York.

\bibitem{gallavotti}
Gallavotti, G. and Cohen, E. G. D. (1995)
Dynamical ensembles in nonequilibrium statistical mechanics.
{\em Phys. Rev. Lett.} {\bf 74}, 2694--2697.


\bibitem{gaspard_book}
Gaspard, P. (1998) {\em Chaos, Scattering and Statistical Mechanics}, 
Cambridge Univ. Press, London.

\bibitem{gaspard}
Gaspard, P. (2004)
Time-reversed dynamical entropy and irreversibility in 
Markovian random processes.
{\em J. Stat. Phys.}  {\bf 117}, 599--615.


\bibitem{gq-pre-10}
Ge, H. and Qian, H. (2010)
The physical origins of entropy production, free energy dissipation 
and their mathematical representations.
{\em Phys. Rev. E} {\bf 81}, 051133.

\bibitem{gq-pre-16}
Ge, H. and Qian, H. (2016)
Mesoscopic kinetic basis of macroscopic chemical thermodynamics: 
A mathematical theory. 
{\em Phys. Rev. E} {\bf 94}, 052150.

\bibitem{gq-jsp-17}
Ge, H. and Qian, H. (2017)
Mathematical formalism of nonequilibrium thermodynamics for nonlinear 
chemical reaction systems with general rate law. 
{\em J. Stat. Phys.}  {\bf 166}, 190--209.


\bibitem{gillespie} 
Gillespie, D. T. (2000) 
The chemical {L}angevin equation. 
{\em J. Chem. Phys.} {\bf 113},  297--306. 

\bibitem{hamermesh}
Hamermesh, M. (1989) {\em Group Theory and Its Application to Physical Problems}, Dover, New York.

\bibitem{HS01}
Hatano, T. and Sasa, S.-I. (2001)
Steady-state thermodynamics of Langevin systems. 
{\em Phys. Rev. Lett.} {\bf 86}, 3463--3466.

\bibitem{jqq}
Jiang, D.-Q., Qian, M. and Qian, M.-P. (2004)
{\em Mathematical Theory of Nonequilibrium Steady States}, 
LNM vol. 1833, Springer, New York.

\bibitem{keizer-book}
Keizer, J. (1987)
{\em Statistical Thermodynamics of Nonequilibrium 
Processes}, 
Springer-Verlag, New York.

\bibitem{Khasminskii12}
Khasminskii, R. (2012) 
{\em Stochastic stability of differential equations}. With contributions by G. N. Milstein and M. B. Nevelson. Completely revised and enlarged second edition. Stochastic Modelling and Applied Probability, 66. Springer, Heidelberg.

\bibitem{kurtz}
Kurtz, T. G. (1981) The central limit theorem for Markov chains.  {\em Ann. Probab.}  {\bf 9}, 557--560.

\bibitem{lebowitz99}
Lebowitz, J. L. (1999)
Microscipic origins of irreversible macroscopic behavior.
{\em Physica A} {\bf 263}, 516--527.

\bibitem{liao-wang}
Liao, G. and Wang, S. (2018) 
Ruelle inequality of folding type for maps. {\em Math. Z.} {\bf 290}, 509--519.

\bibitem{liu}
Liu, D.-P. (2003)
Ruelle inequality relating entropy, folding entropy, and negative Lyaponov exponents.
{\em Comm. Math. Phys.} {\bf 240}, 531--538.

\bibitem{luo}
Luo, J.-L., Van den Broeck, C. and Nicolis, G. (1984)
Stability criteria and fluctuations around nonequilibrium states.
{\em Zeit. Physik. B} {\bf 56}, 165--170.

\bibitem{mackey_rmp}
Mackey, M. C. (1989)
The dynamic origin of increasing entropy.
{\em Rev. Mod. Phys.} {\bf 61}, 981--1015.

\bibitem{mqw24}
Miao, B., Qian, H. and Wu, Y.-S.
(2024) Emergence of Newtonian deterministic causality from stochastic motions in continuous space and time. arXiv:2406.02405.

\bibitem{Mikami1990} 
Mikami, T. (1990) 
Asymptotic analysis of invariant density of randomly perturbed dynamical systems. {\em Ann. Probab.} {\bf 18}, 524--536.

\bibitem{Onsager-Machlup-1953} 
Onsager, L. and Machlup, S. (1953) 
Fluctuations and irreversible processes. {\em Phys. Rev. } {\bf 91}, 1505--1512.

\bibitem{Pavliotis2014}
Pavliotis, G. A. (2014) 
{\em Stochastic Processes and Applications. Diffusion Processes, the Fokker-Planck and Langevin Equations}, Texts in Applied Mathematics, vol. 60. Springer, New York.

\bibitem{prigogine}
Prigogine, I. (1947)
{\em Etude Thermodynamique des Ph\'{e}nom\`{e}nes Irr\'{e}versibles},
Dunod, Paris.


\bibitem{qian-prsa-01}
Qian, H. (2001)
Mathematical formalism for isothermal linear irreversibility. 
{\em Proc. Roy. Soc. A}  {\bf 457}, 1645--1655.

\bibitem{qian-eec-01}
Qian, H. (2001)
Mesoscopic nonequilibrium thermodynamics of single 
macromolecules and dynamic entropy-energy compensation. 
{\em Phys. Rev. E} {\bf 65}, 016102.

\bibitem{qian_epjst}
Qian, H. (2015)
Thermodynamics of the general diffusion process: {E}quilibrium supercurrent and nonequilibrium driven circulation with dissipation.
{\em Eur. Phys. J. Spec. Top.} {\bf 224},
781--799.

\bibitem{qian-book}
Qian, H. (2019)
Nonlinear stochastic dynamics of complex systems, I:
Chemical reaction kinetic perspective with mesoscopic nonequilibrium thermodynamics.
In {\em Complex Science: An Introduction},
Peletier, M. A., van Santen, R. A. and Steur, E. eds.,
World Scientific, Singapore, pp. 347--373.

\bibitem{qatw_16}
Qian, H., Ao, P., Tu, Y. and Wang, J. (2016) A framework towards understanding mesoscopic phenomena: Emergent unpredictability, symmetry breaking and dynamics across scales. {\em Chem. Phys. Lett.} {\bf 665}, 153--161.

\bibitem{qian-kinematics}
Qian, H., Cheng, Y.-C.  and Yang, Y.-J. (2020)
Kinematic basis of emergent energetics of complex dynamics.
{\em EPL} {\bf 131}, 50002.

\bibitem{qian-jjh}
Qian, H. and Hopfield, J. J. (1996)
Entropy-enthalpy compensation: Perturbation and relaxation 
in thermodynamic systems. 
{\em J. Chem. Phys.} {\bf 105}, 9292--9296.

\bibitem{qkkb-review}
Qian, H., Kjelstrup, S., Kolomeisky, A. B. and Bedeaux, D. (2016)
Entropy production in mesoscopic stochastic thermodynamics 
- Nonequilibrium kinetic cycles driven by chemical potentials, temperatures, and mechanical forces (Topical review).
{\em J. Phys. Cond. Matt.} {\bf 28}, 153004.

\bibitem{qqt-jsp}
Qian, H., Qian, M. and Tang, X. (2002)
Thermodynamics of the general diffusion process:
Time-reversibility and entropy production.
{\em J. Stat. Phys.} {\bf 107}, 1129--1141.


\bibitem{qwy-pams-19}
Qian, H., Wang, S. and Yi, Y. (2019)
Entropy productions in dissipative systems.
{\em Proc. Am. Math. Soc.} {\bf 147}, 5209--5225.

\bibitem{qxz}
Qian, M., Xie, J.-S. and Zhu, S. (2009)
{\em Smooth Ergodic Theory for Endomorphisms}, 
LNM vol. 1978, Springer, New York.

\bibitem{ramshaw}
Ramshaw, J. D. (1986)
Remarks on entropy and irreversibility in non-Hamiltonian system.
{\em Phys. Lett. A} {\bf 116}, 110--114.

\bibitem{ruelle_book}
Ruelle, D. (2004) {\em Thermodynamic Formalism: The Mathematical Structure of Equilibrium Statistical Mechanics}. 2$^{nd}$  ed., Cambridge Univ. Press, London. 

\bibitem{ruelle}
Ruelle, D. (1996)
Positivity of entropy production in nonequilibrium statistical 
mechanics. 
{\em J. Stat. Phys.} {\bf 85}, 1--23. 

\bibitem{ruelle99}
Ruelle, D. (1999)
Gaps and new ideas in our understanding of nonequilibrium.
{\em Physica A} {\bf 263}, 540--544.

\bibitem{SFEF24}
Santolin, D., Freitas, N., Esposito, M. and Falasco, G. (2024)
Bridging Freidlin-Wentzell large deviations theory and stochastic thermodynamics. arXiv:2409.07599.

\bibitem{schellman}
Schellman, J. A. (1987)
The thermodynamic stability of proteins.
{\em Annu. Rev. Biophys. Biophys. Chem.}
{\bf 16}, 115--137.

\bibitem{schnakenberg}
Schnakenberg, J. (1976)
Network theory of microscopic and macroscopic behavior 
of master equation.
{\em Rev. Mod. Phys.} {\bf 48}, 571--585.

\bibitem{seifert-review}
Seifert, U. (2012)
Stochastic thermodynamics, fluctuation theorems, and 
molecular machines.
{\em Rep. Progr. Phys.} {\bf 75}, 126001.

\bibitem{Sheu84}
Sheu, S.-J. (1984)
Asymptotic behavior of transition density of diffusion Markov process with small diffusion.
{\em Stochastics} {\bf 13}, 131--163.

\bibitem{Sheu1986}
Sheu, S.-J. (1986)
Asymptotic behavior of the invariant density of a diffusion Markov process with small diffusion. {\em SIAM J. Math. Anal.} {\bf 17}, 451--460.

\bibitem{SF12}
Spinney, R. E.  and Ford, I. J. (2012) 
Entropy production in full phase space for continuous stochastic dynamics. 
{\em Phys. Rev. E} {\bf 85}, 051113.


\bibitem{VE2010} 
Van den Broeck, C. and Esposito, M. (2010)
Three faces of the second law. II. Fokker-Planck formulation. \emph{Phys. Rev. E} {\bf 82}, 011144.


\bibitem{walters}
Walters, P. (1982) 
{\em An Introduction to Ergodic Theory}. Springer-Verlag, New York.

\bibitem{wq-jsp-2020}
Wang, Y. and Qian, H. (2020)
Mathematical representation of Clausius' and Kelvin's statements 
of the second law and irreversibility.
 {\em J. Stat. Phys.} {\bf 179}, 808--837.


\bibitem{cnyang}
Yang, C. N. (1996)
Symmetry and physics.
{\em Proc. Am. Philos. Soc.}
{\bf 140}, 267--288.

\bibitem{young1}
Young, L.-S. (2002)
What are SRB measures, and which dynamical systems have them?
{\em J. Stat. Phys.}  {\bf 108}, 733--754.


\end{thebibliography}
\end{document}